\documentclass[]{spie}   
\usepackage{graphicx}
\usepackage{subfigure,amsmath}
\usepackage[suffix=]{epstopdf}

\newcommand{\C}{\mathcal C}
\newcommand{\eye}{\mathcal I}
\newcommand{\e}[1]{\times 10^{#1}} 

\title{Broadband Focal Plane Wavefront Control of Amplitude and Phase Aberrations}
\author{Tyler D. Groff\supit{a}, N. Jeremy Kasdin\supit{a}, Alexis Carlotti\supit{a} and A J Eldorado Riggs\supit{a}
\skiplinehalf
\supit{a}Princeton University, Princeton, NJ USA
}

\authorinfo{Send correspondence to Tyler Groff \texttt{tgroff@princeton.edu}}

\begin{document}
\maketitle

\begin{abstract} 
The Stroke Minimization algorithm developed at the Princeton High Contrast Imaging Laboratory  has proven symmetric dark hole generation using minimal stroke on two deformable mirrors (DM) in series. The windowed approach to Stroke Minimization has proven symmetric dark holes over small bandwidths by using three wavelengths to define the bandwidth of correction in the optimization problem. We address the relationship of amplitude and phase aberrations with wavelength, how this changes with multiple DMs, and the implications for simultaneously correcting both to achieve symmetric dark holes. Operating Stroke Minimization in the windowed configuration requires multiple wavelength estimates. To save on exposures, a single estimate is extrapolated to bounding wavelengths using the established relationship in wavelength to produce multiple estimates of the image plane electric field. Here we demonstrate better performance by improving this extrapolation of the estimate to other wavelengths. The accuracy of the functional relationship will ultimately bound the achievable bandwidth, therefore as a metric these results are also compared to estimating each wavelength separately. In addition to these algorithm improvements, we also discuss a laboratory upgrade and how it can better simulate broadband starlight. We also discuss the possibility of leveraging two DMs in series to directly estimate the electric field over a narrow bandwidth and the challenges associated with it. 
\end{abstract}


\section{Introduction}\label{sec:intro}
There has been much research into space-based missions for direct imaging of extrasolar terrestrial planets. Two approaches that have been proposed for direct imaging in visible to near-infrared light are occulters, which utilize large external screens to block the starlight while allowing the planet light to pass, and coronagraphs, which use internal masks and stops to change the point spread function of the telescope, creating regions in the image of high contrast where a dim planet can be seen. Observing over a wide band is challenging in both cases and coronagraphs in particular have an extreme sensitivity  to wavefront aberrations generated by the errors in the system optics (occulters are immune to this problem because the starlight never enters the telescope). The chromatic effect of these errors will manifest as aberrations at the image plane that will scale differently in wavelength depending on the type of error and it's location relative to the pupil plane. This necessitates wavefront control algorithms with multi-DM control to correct for the aberrations and allow relaxation of manufacturing tolerances and stability requirements within the observatory to achieve symmetric dark holes in broadband light. In this paper we discuss the challenges associated with wavefront estimation and control, both monochromatic and broadband, in a coronagraphic imager.  We also note that many of these challenges are shared by both space and ground coronagraphs.

\section{Experimental Setup: Princeton High Contrast Imaging Laboratory}\label{HCIL}
\begin{figure}[ht!]
\centering
\includegraphics[width = 0.6\textwidth]{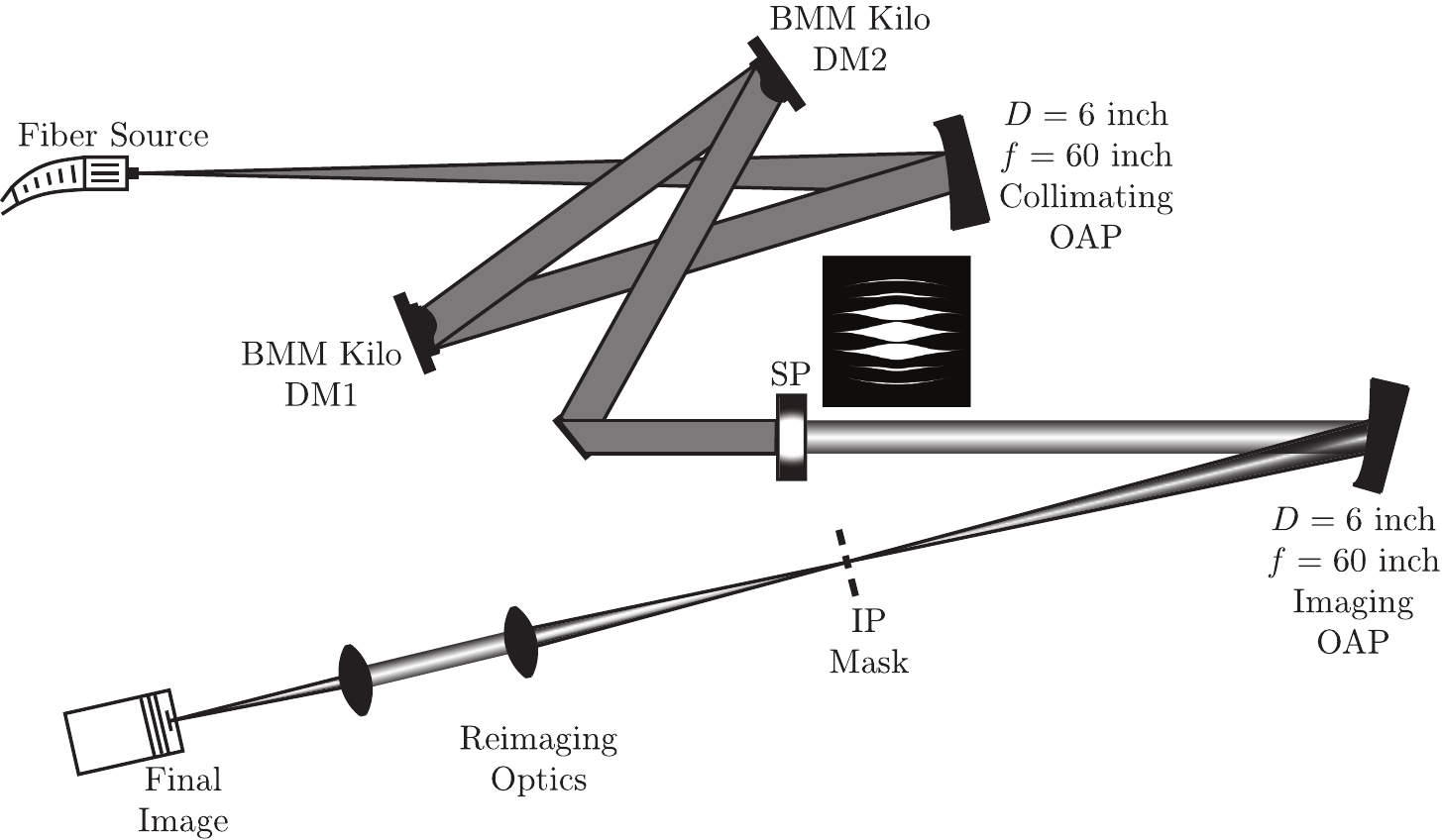}
\caption[HCIL Optical Layout]{Optical ayout of the Princeton HCIL.  Collimated light is incident on two DMs in series, which propagates through a Shaped Pupil, the core of the PSF is removed with an image plane mask, and the $90^\circ$ search areas are reimaged on the final camera.}
\label{fig:layout}
\end{figure}

The High Contrast Imaging Laboratory (HCIL) at Princeton tests coronagraphs and wavefront control algorithms for quasi-static speckle suppression. The collimating optic is a six inch off-axis parabola (OAP) followed by two first generation Boston Micromachines kilo-DMs in series and a shaped pupil coronagraph, which is imaged with a second six inch OAP (Fig.~\ref{fig:layout}).  
\begin{figure}[ht]
\centering
\subfigure[Shaped Pupil] {
    \label{SP}
    \includegraphics[width = .205\textwidth]{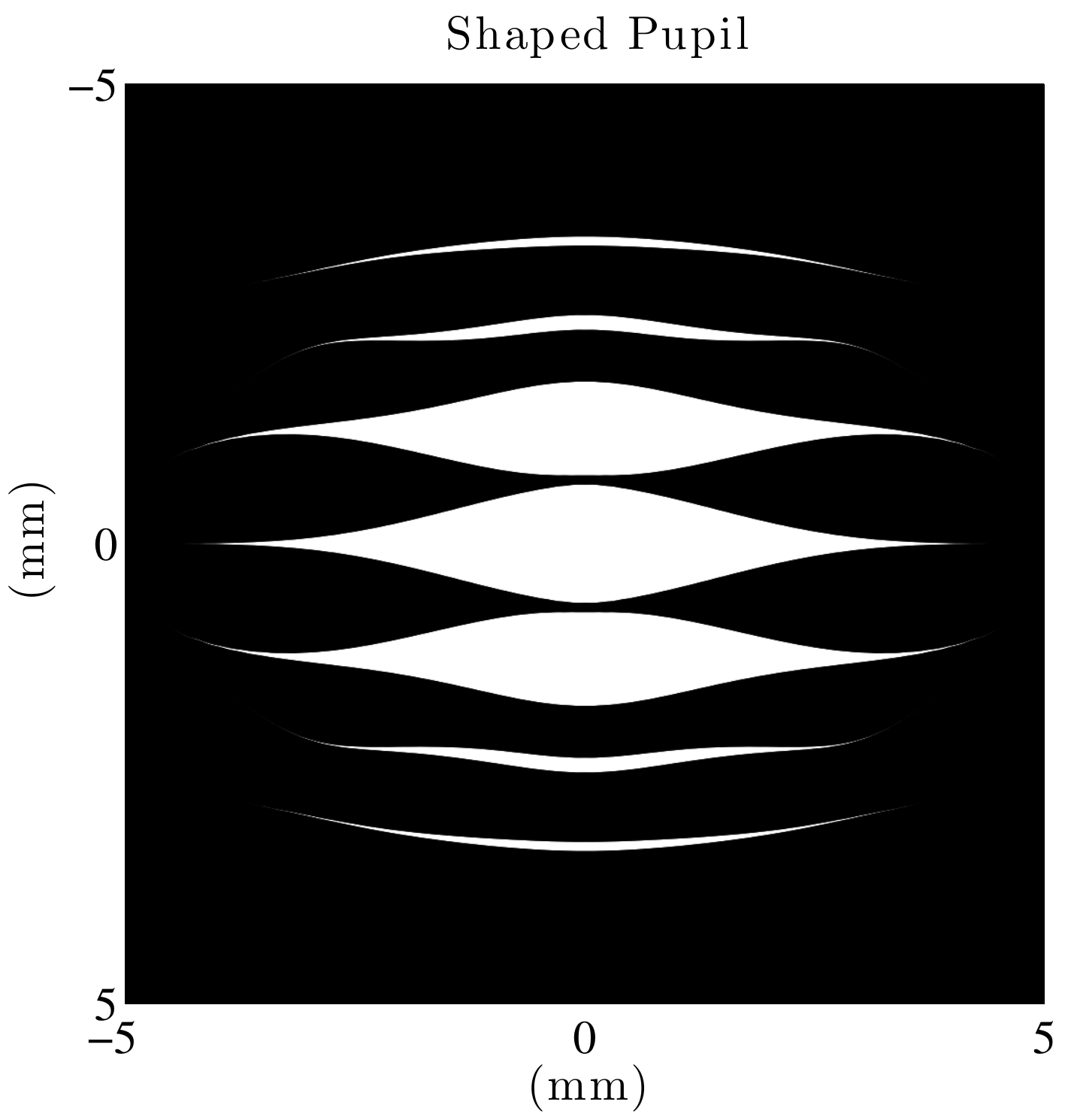}}
\subfigure[Ideal PSF] {
    \label{Ideal_PSF}
    \includegraphics[width = .24\textwidth]{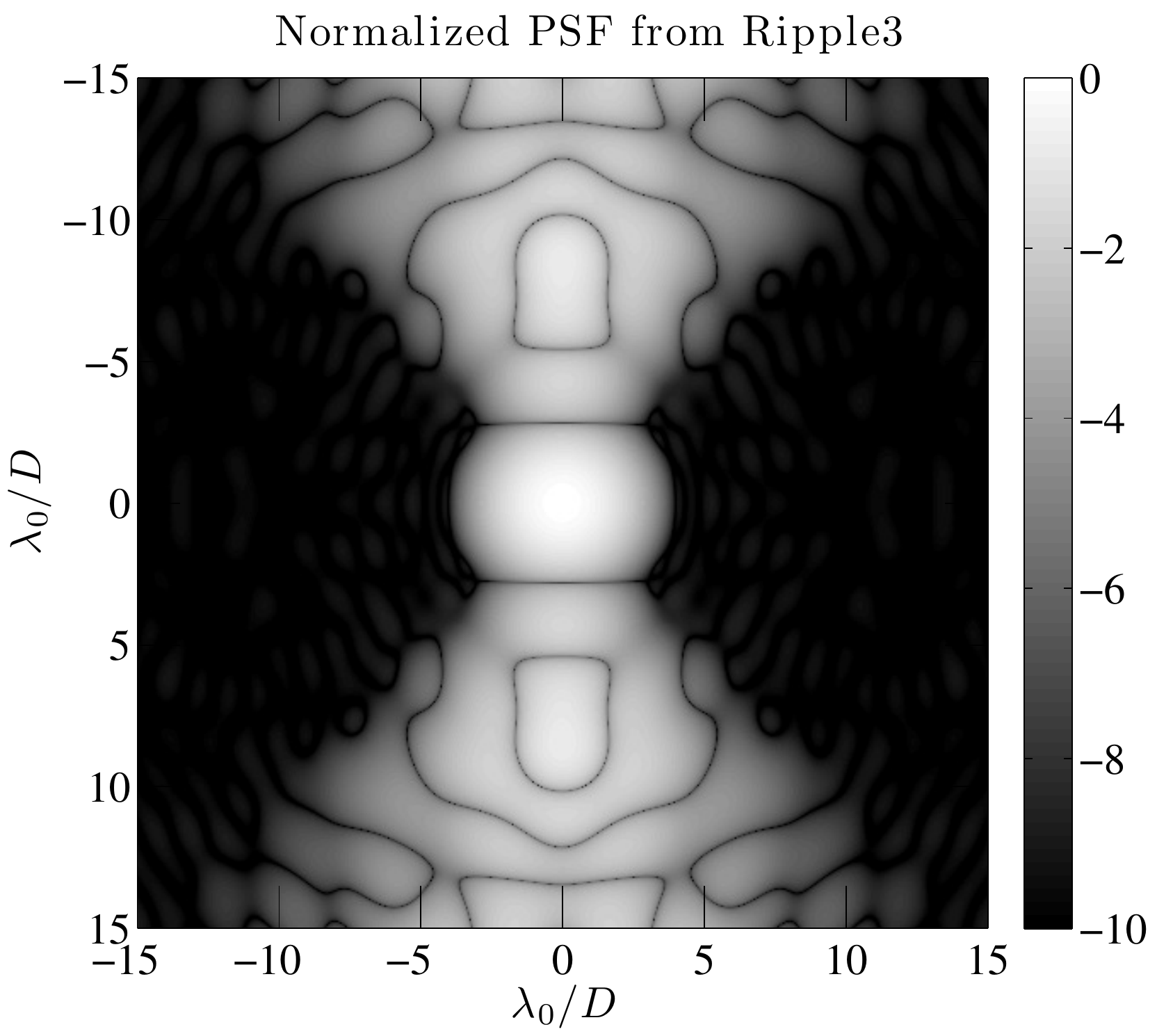}}
   \subfigure[Aberrated Pupil]{
    \label{ab_SP}
    \includegraphics[width = .205\textwidth]{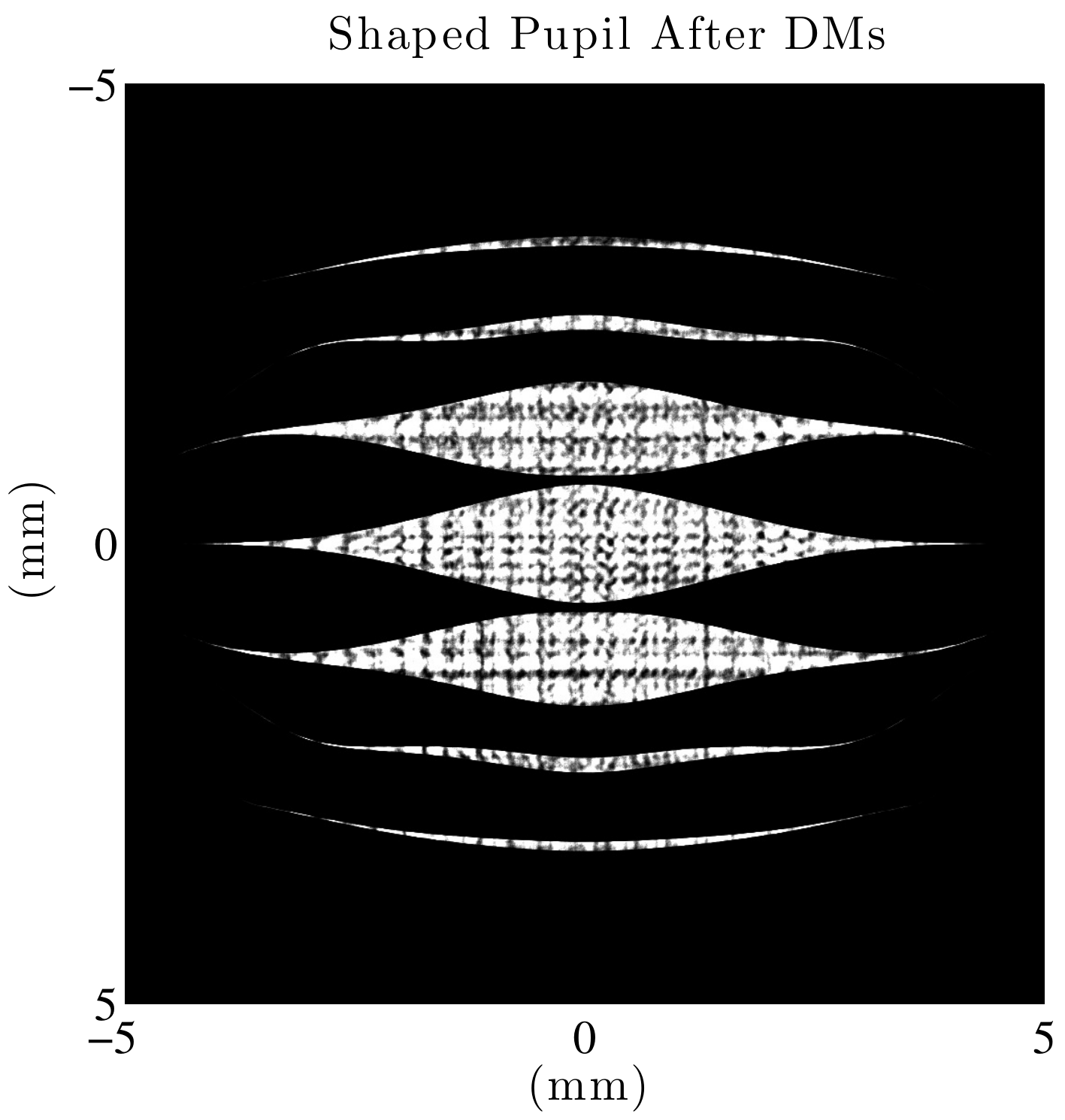}\label{SPab}}
    \subfigure[Aberrated PSF]{
    \label{ab_PSF}
    \includegraphics[width = .24\textwidth]{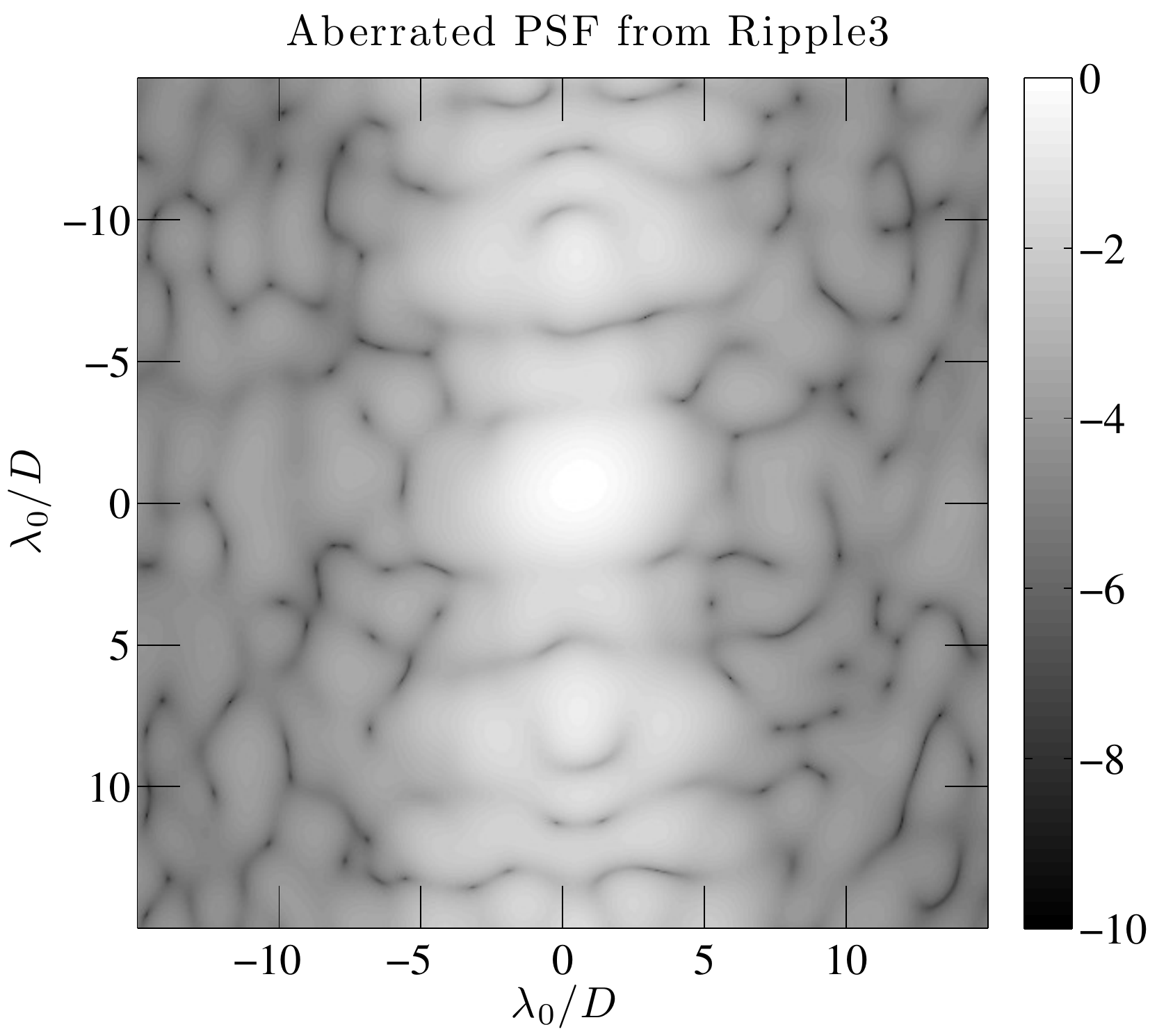}}
    \caption[Ideal vs. Aberrated PSF]{Example of the effect of an aberrated field incident on a Shaped Pupil coronagraph. The aberrations are simulated by Fresnel propagating the measured nominal shapes of the HCIL DMs to the pupil plane. Other sources of aberrations are not included because they have not been measured. (d) The PSF of the shaped pupil with the simulated aberrations.   The figures are in a log scale, and the log of contrast is shown in the colorbars.}
    \label{fig:expup}
\end{figure}
We use a shaped pupil coronagraph, shown in Fig.~\ref{SP}, and described in detail in Belikov et al.\cite{belikov2007broadband}.  This coronagraph produces a discovery space with a theoretical contrast of $3.3\times10^{-10}$ in two $90^\circ$ regions as shown in Fig.~\ref{Ideal_PSF}. At the Princeton HCIL, the  aberrations in the system result in an uncorrected average contrast of approximately $1\times10^{-4}$ in the area immediately surrounding the core of the point spread function (PSF), which agrees with the simulations shown in Fig.~\ref{ab_PSF}. Since the coronagraph is a binary mask, its contrast performance is fundamentally achromatic, subject only to the physical scaling of the PSF with wavelength. The lab can be configured with either a  $635$ nm monochromatic laser diode input, or a Koheras supercontinuum source.
\begin{figure}[ht]
\centering
\includegraphics[width = .5\textwidth]{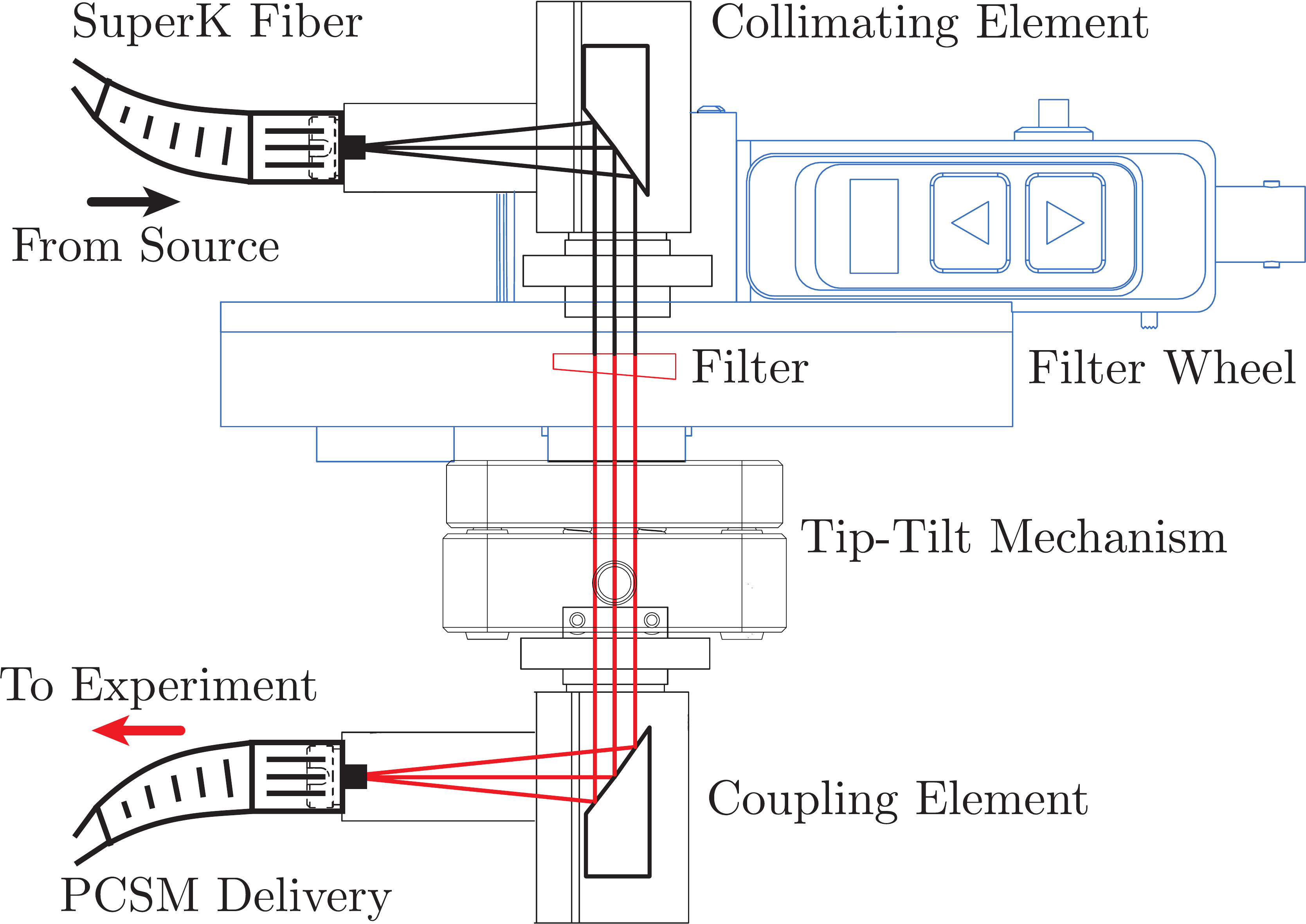}
\caption[HCIL Filter Mechanism]{Optical Layout of the Princeton HCIL's Filter Wheel. The light from the SuperK supercontinuum fiber is collimated by a Thorlabs reflective collimator (c, passes through a filter wheel which contains narrow band interference filter, and is recoupled into a Koheras Photonic Crystal continuously Single Mode (PCSM) fiber with another reflective coupler (RC08FC). The system is rigid with the exception of the tip-tilt mechanism, which is used to align the beam for coupling back into the PCSM fiber that delivers light onto the bench.}
\label{fig:filterassy}
\end{figure}
As shown in Fig.~\ref{fig:filterassy}, before the super continuum source is injected into the laboratory experiment, it is first collimated by a $90^\circ$ off axis parabolic element designed specifically for collimating/coupling of polychromatic light from a fiber. After the light is collimated it passes through a filter wheel where a set of interference filters allows us to sample narrow bandwidths in a $20\%$ range around $635$ nm. After the light passes through the filter wheel it is re-coupled with a second off axis element into a second fiber made by Koheras which is designed to be continuously single mode over the entire visible and near-infrared spectrum. This allows us to better reproduce the wavelength nature of a light coming from a star, and solves the problem of multimode output at shorter wavelengths (as well as high attenuation at longer wavelengths). Since the collimating/coupling elements rigidly attach the fiber tips to the $90^\circ$ OAPs, alignment of the beam is determined entirely by tip-tilt variation of the collimated beam. To precisely recouple the light back into the delivery fiber, the collimating element is rigidly mounted to the filter wheel and the coupling element is mounted to a tip-tilt stage. To eliminate ghosts, all interference filters have a small wedge between their exterior surfaces. To guarantee a quality alignment for all of the filters for a fixed tip-tilt, they must all be clocked inside the filter wheel so that when they are positioned within the beam, the wedge is aligned in the same direction. To guarantee stability of the coupling (which is sensitive on the sub-micron level) the entire optical train is sealed from the outside environment, eliminating any air flow through the system. With the system very compact and light, sealed, and highly rigid (since the tip-tilt mechanism is very stiff) we observe that the coupling is reliable over a period of weeks to months once it has been aligned. Since the original HCIL experiment had proved to be limited by the stability of its old HeNe laser and its free space coupling into a fiber this was a critical design parameter for the filtering scheme.

The two source configurations allows us to test control algorithms in both monochromatic and broadband light (typically $\sim 10-20\%$ of the central wavelength).The monochromatic experiments allow us to test controller performance very quickly, while leaving the results independent of any chromatic effects. Once a particular algorithm has been proven in monochromatic light, we can use the polychromatic configuration to test its performance over a larger bandwidth. 

\section{Monochromatic Correction: Correcting Amplitude and Phase}
In this paper we use the stroke minimization algorithm\cite{pueyo2009optimal} as the controller to test our estimator. It corrects the wavefront by minimizing the actuator stroke on the DMs subject to a target contrast value\cite{pueyo2009optimal}.  Expressing the DM actuator amplitudes as a vector, $u$, the optimization problem can be written as
\begin{equation}
\begin{array}{ccc}
\mbox{minimize} & & \sum_{k=1}^N a_k^2 = u^Tu \medskip\\
\mbox{subject to} & & I_{DZ} \le 10^{-C},
\end{array}\label{eq:coststate}
\end{equation}
where $a_k$ is the commanded height of actuator $k$. The contrast level, $I_{DZ}$, is the residual intensity in the dark hole after the previous correction, and $C$ is the target contrast.  We solve the optimization by approximating $I_{DZ}$ as a quadratic form,

\begin{equation}
I_{DZ} \cong \left(\frac{2\pi}{\lambda_0}\right)^2u^TM_0u + \frac{4\pi}{\lambda_0} \Im \{b_0^T\} u+d_0,
\end{equation}
where $b_0$ is a vector describing the interaction between the DM shape and the aberrated field, $d_0$ is a vector that expresses contrast in the dark hole, and $M_0$ is the matrix which describes the linearized mapping of DM actuation to intensity in the dark hole:
\begin{align}
M_0 &= <\C\{Af\},\C\{Af\}> \label{eq:monom}\\
b_0 &= <\C\{A(1+g)\},\C\{Af\}> \label{eq:monob}\\
d_0 &= <\C\{A(1+g)\},\C\{A(1+g)\}>.
\end{align}
The resulting quadratic subprogram is easily solved by augmenting the cost function via Lagrange multiplier, $\mu_0$, and solving for the commanded actuator heights:
\begin{align}
J &= u^T\left( \eye + \mu_0\frac{4 \pi^2}{\lambda_0^2} M_0\right)u + \mu_0 \frac{4 \pi}{\lambda_0} u^T \Im\{b_0\} + \mu_0 \left(d_0 - 10^{-C}\right). \label{eq:monocost}\\
u_{opt} &= - \mu_0 \left(\frac{\lambda_0}{2\pi} \eye + \mu_0 \frac{2\pi}{\lambda_0} M_0 \right)^{-1} \Im\{b_0\}.\label{eq:monocontrol}
\end{align}

We find the optimal actuator commands via a line search on $\mu$ to minimize the augmented cost function (Eq.~(\ref{eq:monocost})). It is shown in Pueyo et al.\cite{pueyo2009optimal} that this is a quadratic subprogram of the full nonlinear problem, meaning we can iterate to reach any target contrast. In addition to regularizing the problem of minimizing the contrast in the search area, minimizing the stroke has the added advantage of keeping the actuation small and thus within the linear approximation. If the DM model and its transformation to the electric field (embedded in the $M$ matrix) were perfectly known, the achievable monochromatic contrast using stroke minimization would be limited only by estimation error as long as the DM actuation remains within the bounds of the linearization.  Our ability to estimate the field is driven  largely by the residual model error associated with the DM. 
\begin{figure}[h!]
\centering
\subfigure[]{\includegraphics[width = 0.32\textwidth]{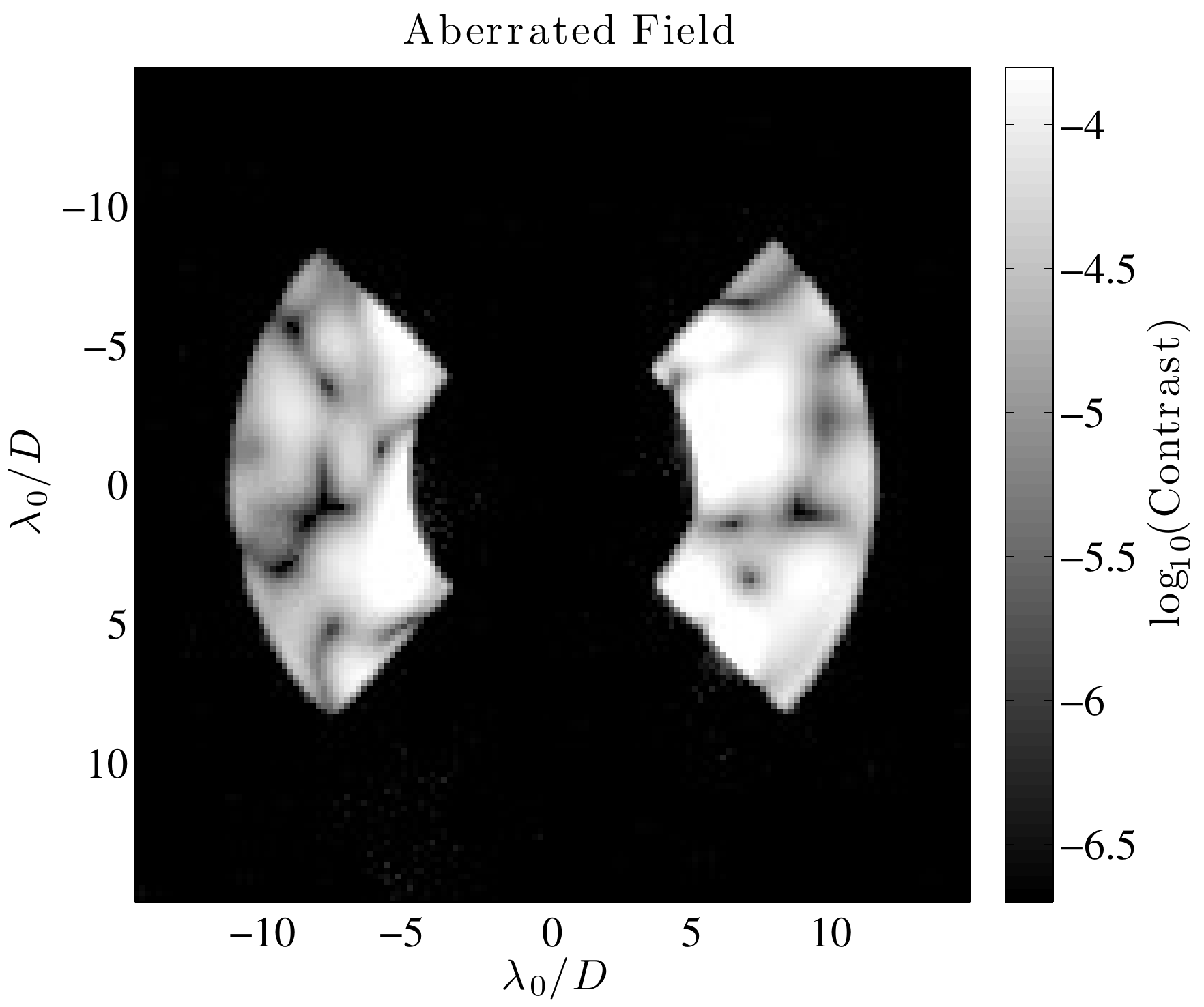}\label{fig:onepair_initial}}
\subfigure[]{\includegraphics[width = 0.335\textwidth]{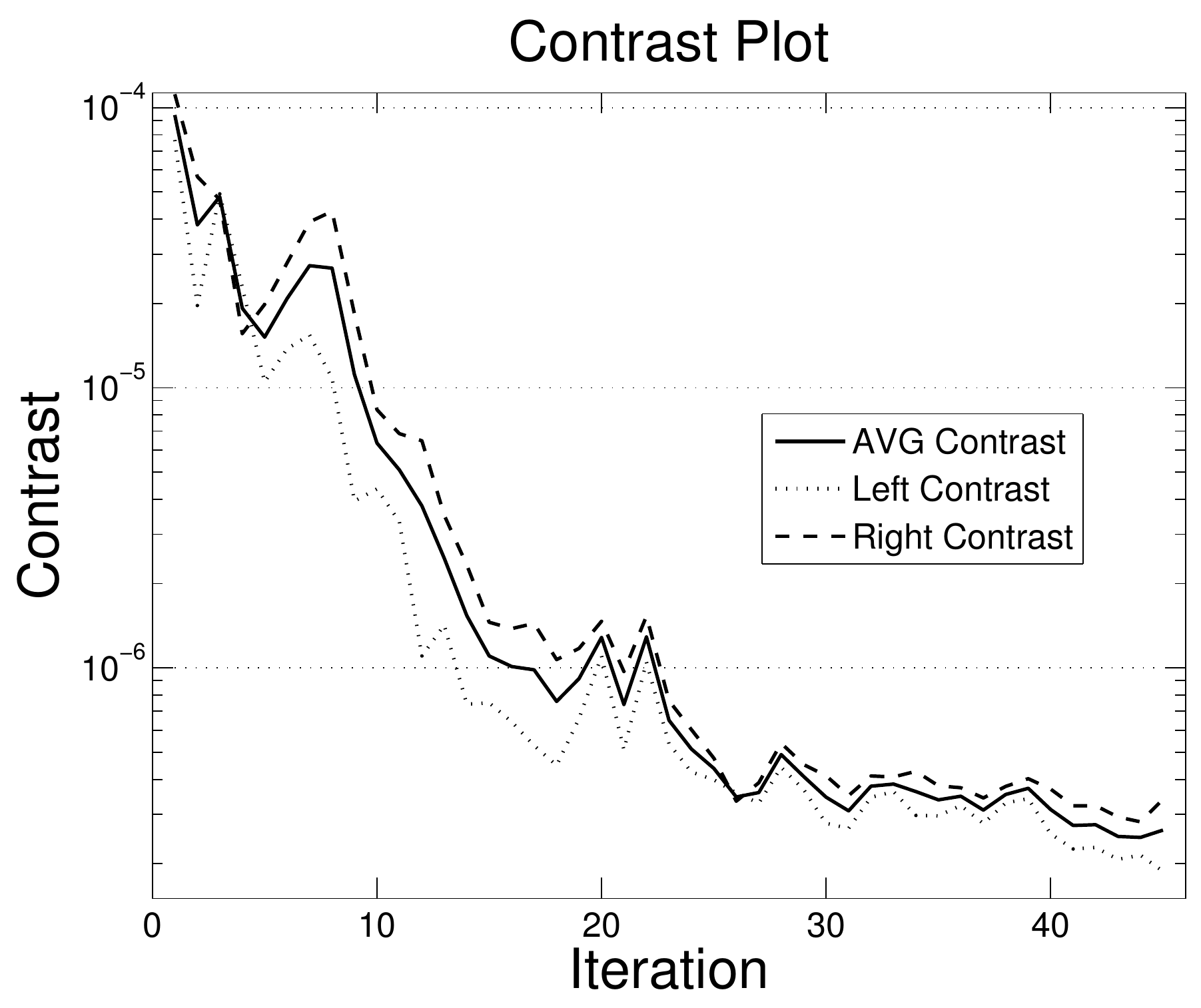}\label{fig:onepair_contrast}}
\subfigure[]{\includegraphics[width = 0.32\textwidth]{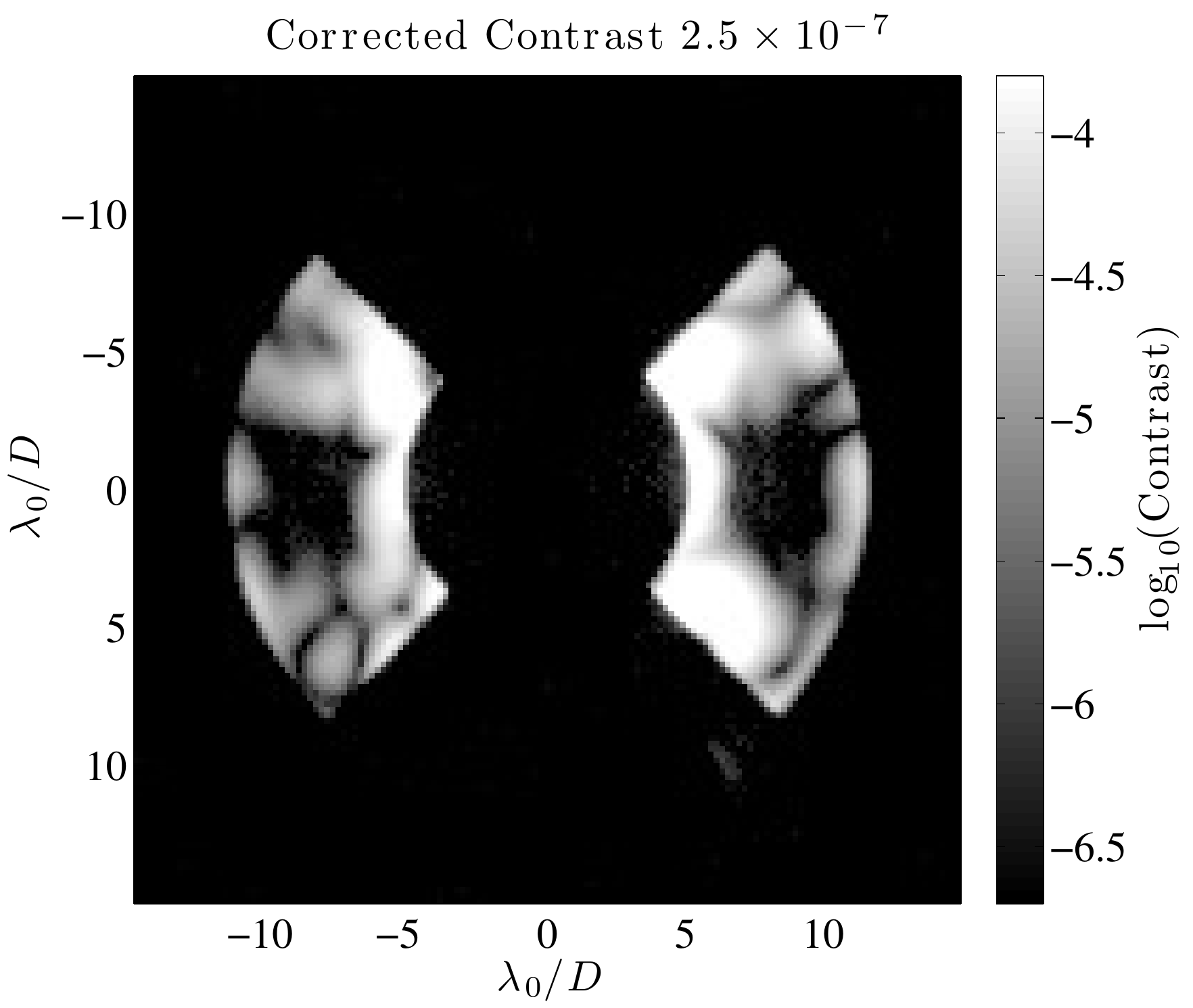}\label{fig:onepair_final}}
\caption{Experimental results of sequential DM correction using the discrete time extended Kalman filter with one image pair to build the image plane measurement, $z_k$.  The dark hole is a square opening from 7--10 $\times$ -2--2 $\lambda/D$ on both sides of the image plane.  (a) The aberrated image.  (b)  Contrast plot. (c) The corrected image. Image units are log(contrast). }\label{fig:onepair}
\end{figure}

\begin{figure}[ht!]
\centering
\subfigure[] {
    \label{monoab}
    \includegraphics[width = 0.32\textwidth,clip=true,trim=.05in 0in .35in 0in]{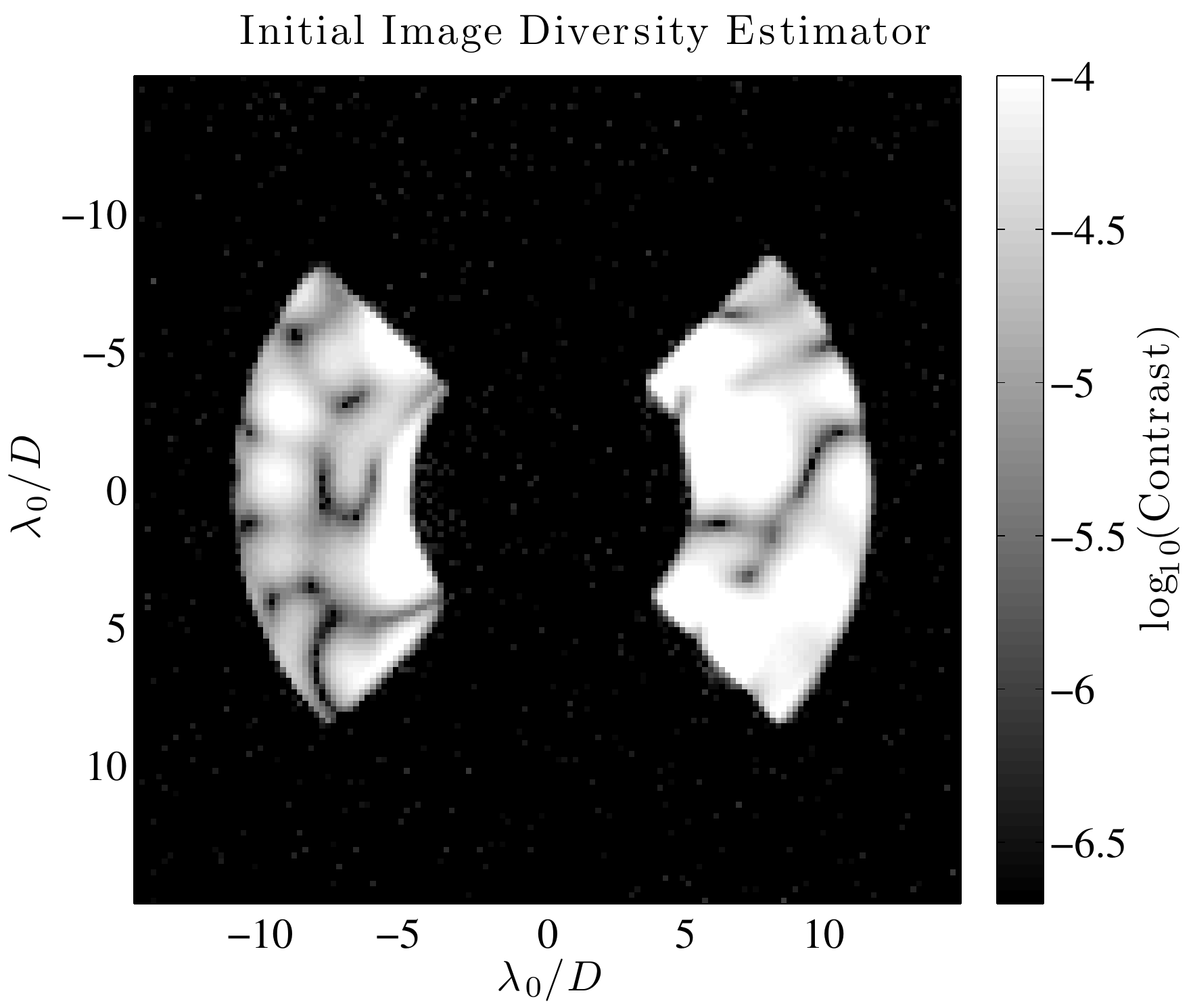}}
\subfigure[] {
    \label{monobest}
    \includegraphics[width = 0.32\textwidth,clip=true,trim=.05in 0in .35in 0in]{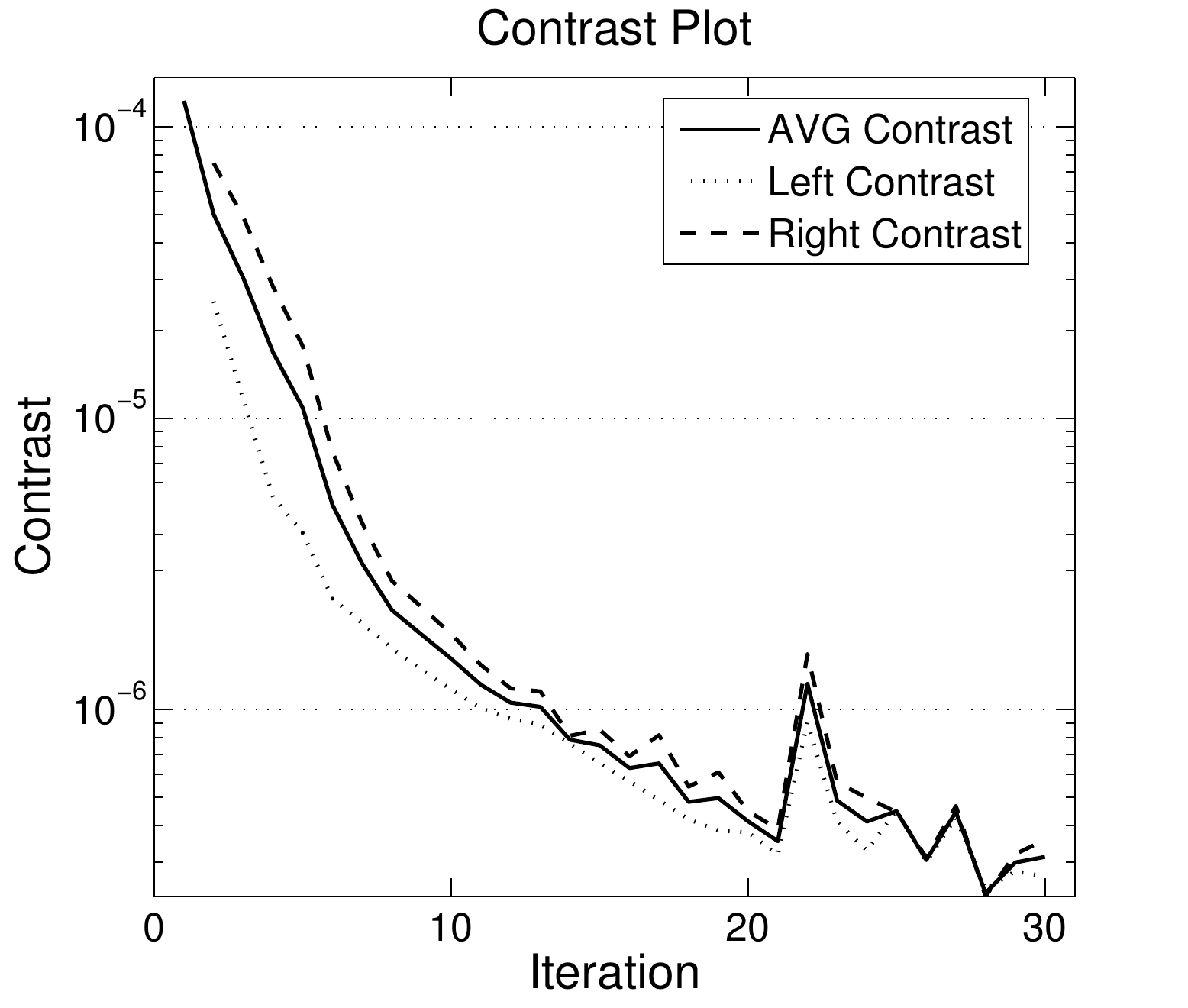}}
    \subfigure[]{
    \label{monoplot}
    \includegraphics[width = 0.32\textwidth,clip=true,trim=.05in 0in .5in 0in]{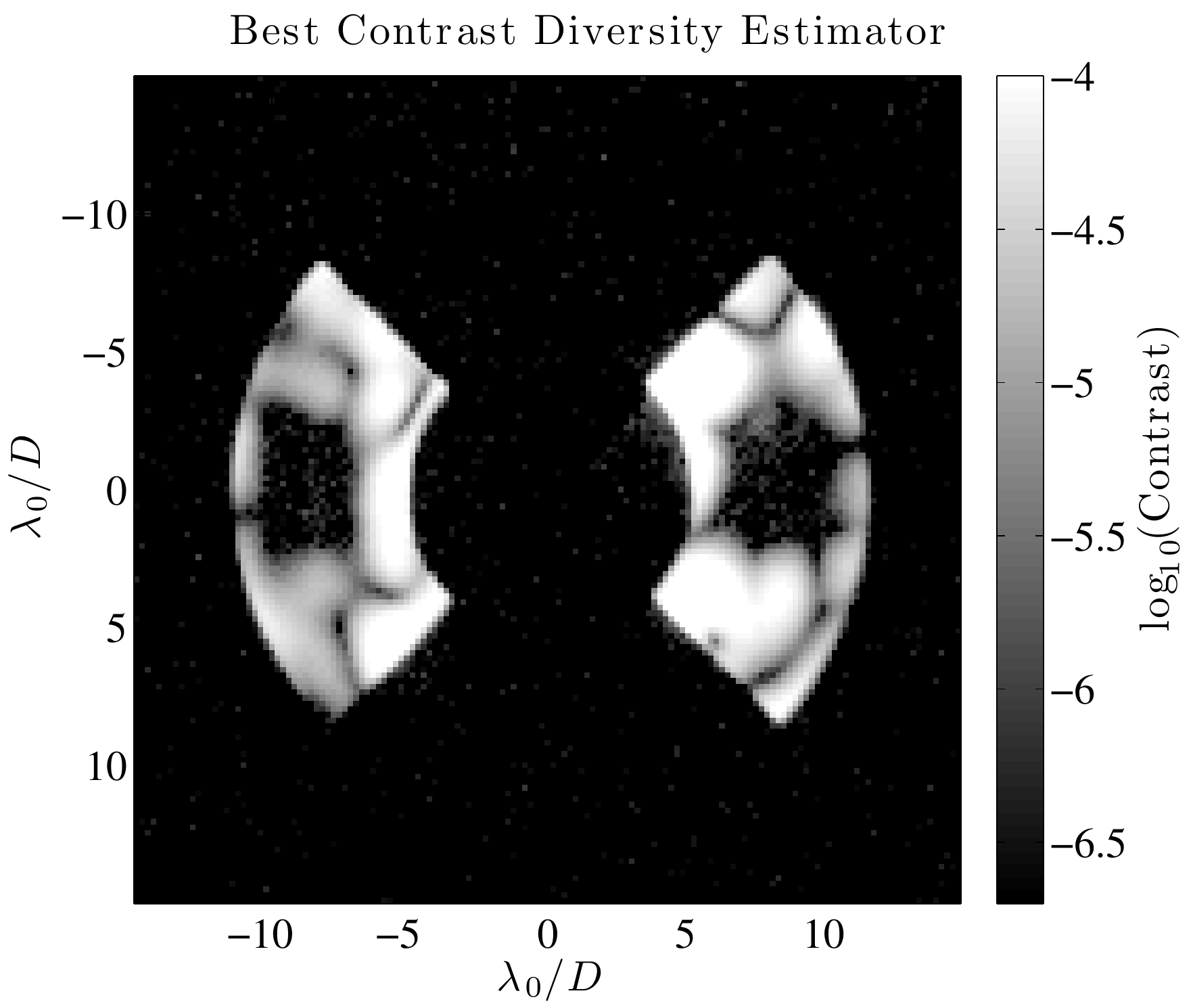}}
    \caption{(a) Uncorrected Image  (b)  Corrected Image, $2.3\times10^{-7}$ Avg. Contrast  (c) Contrast at each iteration}
    \label{fig:mono_contrast}
\end{figure}

The estimation technique that has provided the best results to date over a small area at the Princeton HCIL is the Kalman filter estimation scheme\cite{groff2011designing} using pairwise images\cite{borde2006speckle,giveon2007broadband}. Our ability to estimate the field is driven  largely by the residual  model error associated with the DM and its effect on the image plane electric field. Recent improvements have allowed us to reach an average value of $2.5\times10^{-7}$ in a (7-10)x(-2-2) $\lambda/D$ region (Fig.~\ref{fig:onepair}) on both sides of the image plane with approximately one third of the previously required measurements\cite{groff2012aeroconf,groff2011progress}, a direct result of the Kalman filter estimator. Since the broadband results reported here still use the DM Diversity estimation algorithm\cite{borde2006speckle,giveon2007broadband} to estimate the field, we show these monochromatic results as a baseline for the rest of the paper. Using this estimator, the monochromatic stroke minimization has produced an average contrast of $2.5\times10^{-7}$ in a (7-10)x(-2-2) $\lambda/D$ region (Fig.~\ref{fig:mono_contrast}) on both sides of the image plane, albeit with many more exposures than the Kalman filter results (since there are 4 times as many measurements taken per iteration).

\section{Windowed Stroke Minimization: Broadband Optimization}\label{broad_single}
Narrowband correction schemes ($ \Delta \lambda / \lambda \leq 2\%$) for high contrast imaging have been well demonstrated \cite{trauger2007nature}. Nevertheless, achieving broadband correction is key to raising the technology readiness level (TRL) of wavefront control algorithms.  Generating a null for each wavelength separately to spectrally characterize a target would be prohibitively slow because of the large number of exposures required to estimate the electric field. A broadband algorithm reduces the number of exposures, and hence the time required to spectrally characterize a target. Increasing the bandwidth is also the easiest way to increase the number of photons in an inherently photon limited system, reducing the exposure time required to achieve a planetary detection. Thus, broadening the spectral range of the wavefront correction will improve the overall efficiency of a planet finding mission and will allow for fewer parallel beam paths, making it cheaper and less complex to measure over a broad bandwidth. It has been shown that two DMs in series can be used to correct over a broader range of wavelengths by incorporating wavelength expansions of the aberrated electric field propagation \cite{pueyo2007polychromatic}. 

We will create a broadband form of the stroke minimization algorithm by augmenting the cost function with multiple wavelengths, requiring an independent estimate for each one. Since estimation requires nearly all of the wavefront correction exposures, taking estimates for each wavelength is only marginally more efficient than correcting each wavelength individually. We solve this problem by extrapolating a single monochromatic estimate to higher and lower wavelengths. Assuming that the amplitude errors at the pupil are wavelength independent and the phase distributions scale inversely with wavelength, the pupil plane electric field is given by
\begin{equation}
E_{pup}(u,v,\lambda)\approx A(u,v) e^{i 2 \pi \frac{\lambda_0}{\lambda} \phi_0(u,v)}. \label{simplepupil}
\end{equation}

Given a linear, wavelength dependent, transformation between the pupil and image plane $\mathcal{C}_\lambda$ (e.g. the optical Fourier transform) we can use Eq.~\ref{simplepupil} to describe the electric field estimate at an arbitrary wavelength $\lambda$ as a function of the image plane electric field estimate taken at the original wavelength $\lambda_0$ using the DM-diversity algorithm. With the wavelength dependence only appearing in the phase of the pupil field, the extrapolated electric field estimate from $\lambda_0$ to $\lambda$ is given by
\begin{equation}
E_{est}(x,y,\lambda) = \mathcal{C}_\lambda \left\{\frac{\mathcal{C}^{-1}_{\lambda_0}\{E_{est}(\lambda_0)\}^{\frac{\lambda_0}{\lambda}}}{\left|\mathcal{C}^{-1}_{\lambda_0}\{E_{est}(\lambda_0)\right\}|^{\frac{\lambda_0}{\lambda}-1}}\right\} .\label{extrapolation}
\end{equation}

With the extrapolation technique in hand, we construct the broadband control law. Given multiple constrained wavelengths, the optimization problem from Eq.~\ref{eq:coststate} is rewritten as
\begin{equation}
\begin{array}{ccc}
\mbox{minimize} & & \displaystyle \sum_{k=1}^{N_{act}} a_k^2 = u^Tu \medskip\\
\mbox{subject to:} & & \displaystyle I_{DH}(\lambda_0)  \le 10^{-C_{\lambda_0}}, \medskip\\
	&& \displaystyle I_{DH}(\lambda_1) \le 10^{-C_{\lambda_1}},\medskip\\
	&& \displaystyle I_{DH}(\lambda_2)  \le 10^{-C_{\lambda_2}}\label{threeconstraint}\medskip\\
\mbox{where} && \lambda_1 = \gamma_1 \lambda_0\\
	&& \lambda_2  = \gamma_2 \lambda_0.
\end{array}
\end{equation}
The algorithm now minimizes actuator strokes, $u$,  under the constraint that a particular contrast, $10^{-C_i}$,  be achieved in three separate wavelengths, $\lambda_i$. The cost function takes on the same basic form, but now includes multiple wavelengths:

\begin{align}
J =& u \left[\mathcal{I} + \mu_0 \frac{4\pi^2}{\lambda_0^2}\left( M_{\lambda_0} + \delta_1 \frac{w(\lambda_1)}{\gamma_1^2} M_{\lambda_1}  + \delta_2 \frac{w(\lambda_2)}{\gamma_2^2} M_{\lambda_2}\right) \right]u^T\bigskip\nonumber\\
 &+  \mu_0 \frac{4\pi}{\lambda_0} \left[\Im\{b_{\lambda_0}\} + \delta_1 \frac{w(\lambda_1)}{\gamma_1} \Im\{b_{\lambda_1}\} + \delta_2 \frac{w(\lambda_2)}{\gamma_2} \Im\{b_{\lambda_2}\}\right]u^T\bigskip\nonumber\\
 &+ \mu_0\left[\left(d_{\lambda_0} - 10^{-C_{\lambda_0}}\right) + \delta_1 w(\lambda_1) \left(d_{\lambda_1} - 10^{-C_{\lambda_1}}\right) + \delta_2 w(\lambda_2) \left(d_{\lambda_2} - 10^{-C_{\lambda_2}}\right) \right].
 \end{align}
 
\noindent where $M_\lambda$ describe the effect on the image plane intensity from the DM actuation, $b_\lambda$ is intensity from the interaction of the DMs with the aberrated field, and $d_\lambda$ is intensity of the uncorrected aberrated field. The multipliers $\delta_1$ and $\delta_2$ allow us to parameterize a single Lagrange multiplier in the cost function. In the more general case with three Lagrange multipliers it is possible that the global minimum of the function would not result in constant contrast at each wavelength. This approach to stroke minimization, what we call ``windowed stroke minimization", makes the optimization in wavelength tractable and allows for estimation only at a single wavelength. Extrapolating estimates and optimizing for discrete wavelengths over the target bandwidth reduces the required number of exposures for correcting over a bandwidth. 
\section{Experimental Results}\label{sec:results}
The results in \S\ref{sub:prepcsm} are prior to upgrading the light source in the laboratory with a photonic crystal single mode fiber (LMA-5 from Koheras), and are included for the sake of comparison. \S\ref{sub:pcsm} shows the most current results from the Princeton HCIL producing symmetric dark holes in a targeted $10\%$ band around $\lambda_0 = 633$ nm. In all cases, the results we present are in a dark hole region with dimension 7--10 $\times$ -2--2 $\lambda_0/D$. The contrast measurement is pinned to the central wavelength so that we can evaluate the performance for a fixed sky angle, $\alpha$, defined as $\alpha = \tan^{-1} (n \lambda_0 /D)$. In a $10\%$ band the physical shift is less than one pixel at the HCIL, and the controller corrects an area of 6--11 $\times$ -3--3 $\lambda_0/D$. Thus, we do not have uncontrolled areas leaking light into the dark hole as the wavelength changes, skewing our measurement of the controller performance. 
\subsection{Prior to Single Mode Photonic Crystal Fiber}\label{sub:prepcsm}
In the first broadband experiments at the HCIL, the output fiber was simply a $633$ nm single mode fiber. The correction was performed at $620$ rather than $633$ nm as in the other experiments, partially because of filter availability. In this experiment we have tested the performance of the windowed stroke minimization algorithm over a $10\%$ bandwidth. The estimate for the filters bounding the $10\%$ target bandwidth are computed using the estimate extrapolation technique.
\begin{figure}[h!]
\centering
\subfigure[]{\includegraphics[width = 0.43\textwidth]{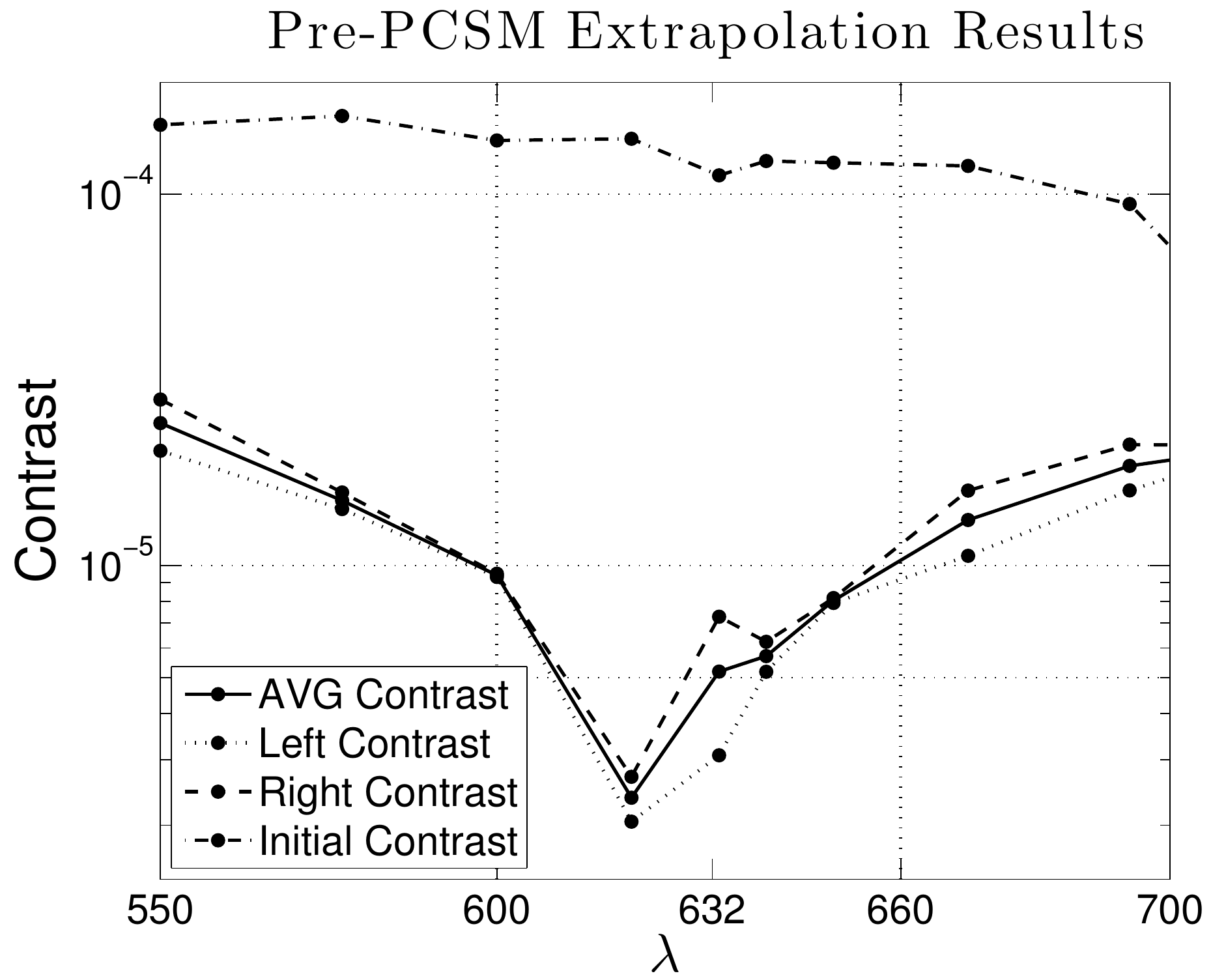}\label{fig:precontrastextrap}}
\subfigure[]{\includegraphics[width = 0.4\textwidth]{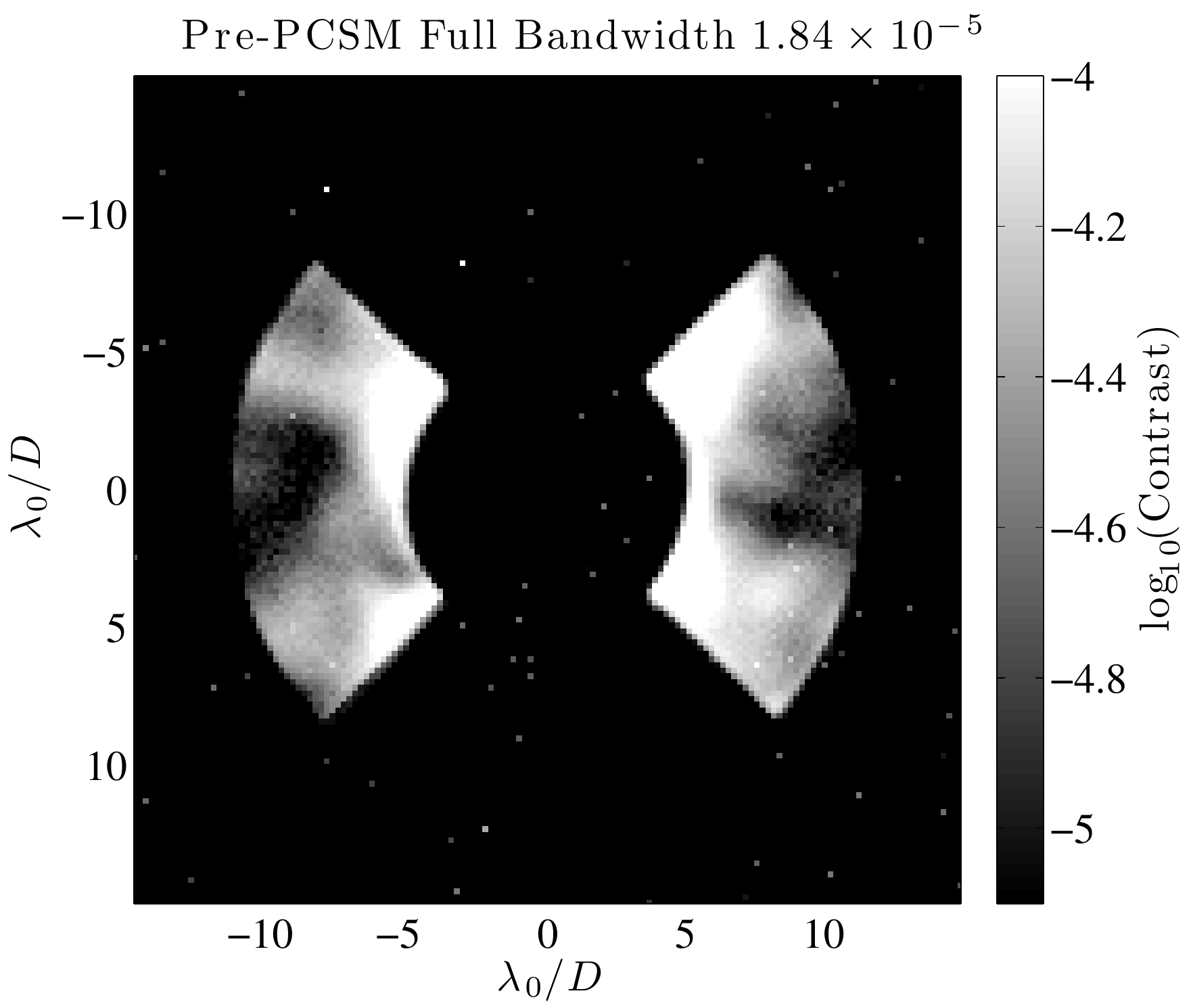}\label{fig:prebroadextrap}}\\
\subfigure[]{\includegraphics[width = 0.4\textwidth]{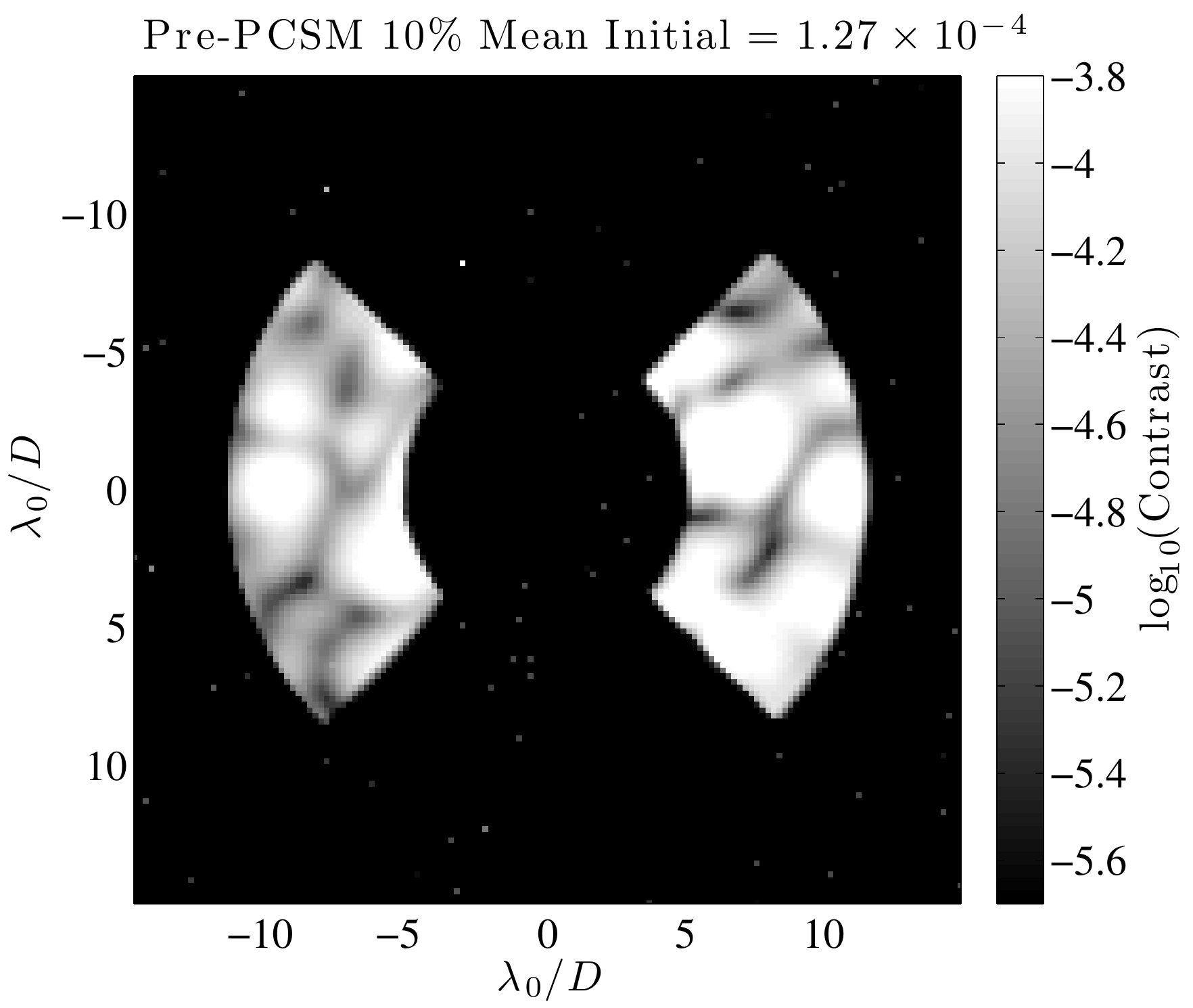}\label{fig:preinitextrap}}
\subfigure[]{\includegraphics[width = 0.4\textwidth]{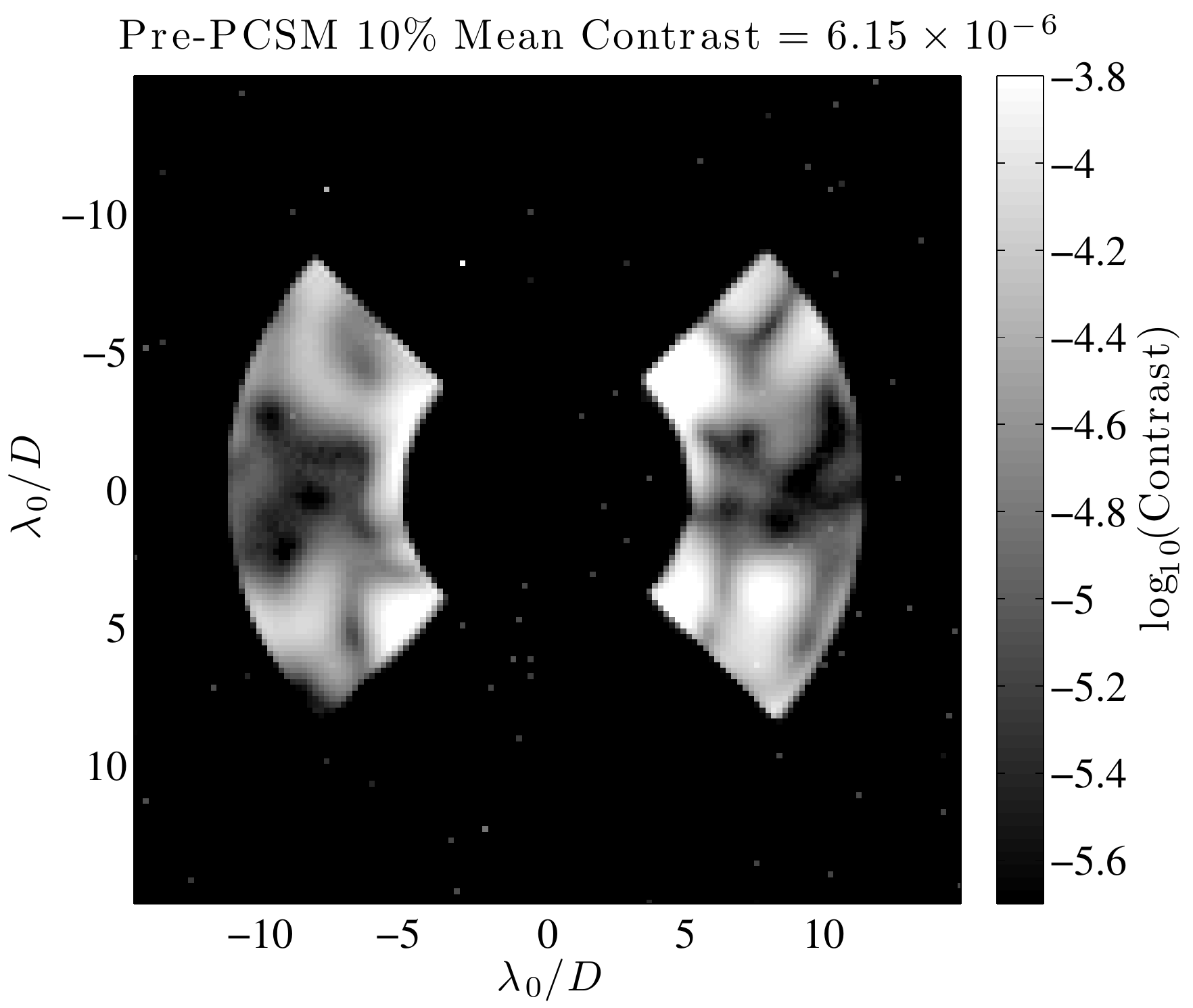}\label{fig:prefinextrap}}
\caption[Pre-PCSM Extrapolated Results]{Pre-PCSM Extrapolated results }\label{fig:preextrap}
\end{figure}
Starting at an average contrast of $1.2740\e{-4}$ (Fig.~\ref{fig:preinitextrap}) over the five filters spanning the $10\%$ bandwidth (600,620,633,640,650 nm), Fig.~\ref{fig:prefinextrap} shows an average contrast of $6.15\e{-6}$ when using the filter extrapolation technique . Note that while the central wavelength of the $650$ nm filter does not exactly reach $5\%$ above our central wavelength, it has a relatively wide bandwidth that reaches $558$ nm at its full-width-half-max. Starting at a contrast level of $1.0529\e{-4}$ over the full bandwidth, the controller reached a contrast limit of $1.842\e{-5}$.
\begin{figure}[h!]
\centering
\subfigure[$\lambda = 600$ nm]{\includegraphics[width = 0.325\textwidth]{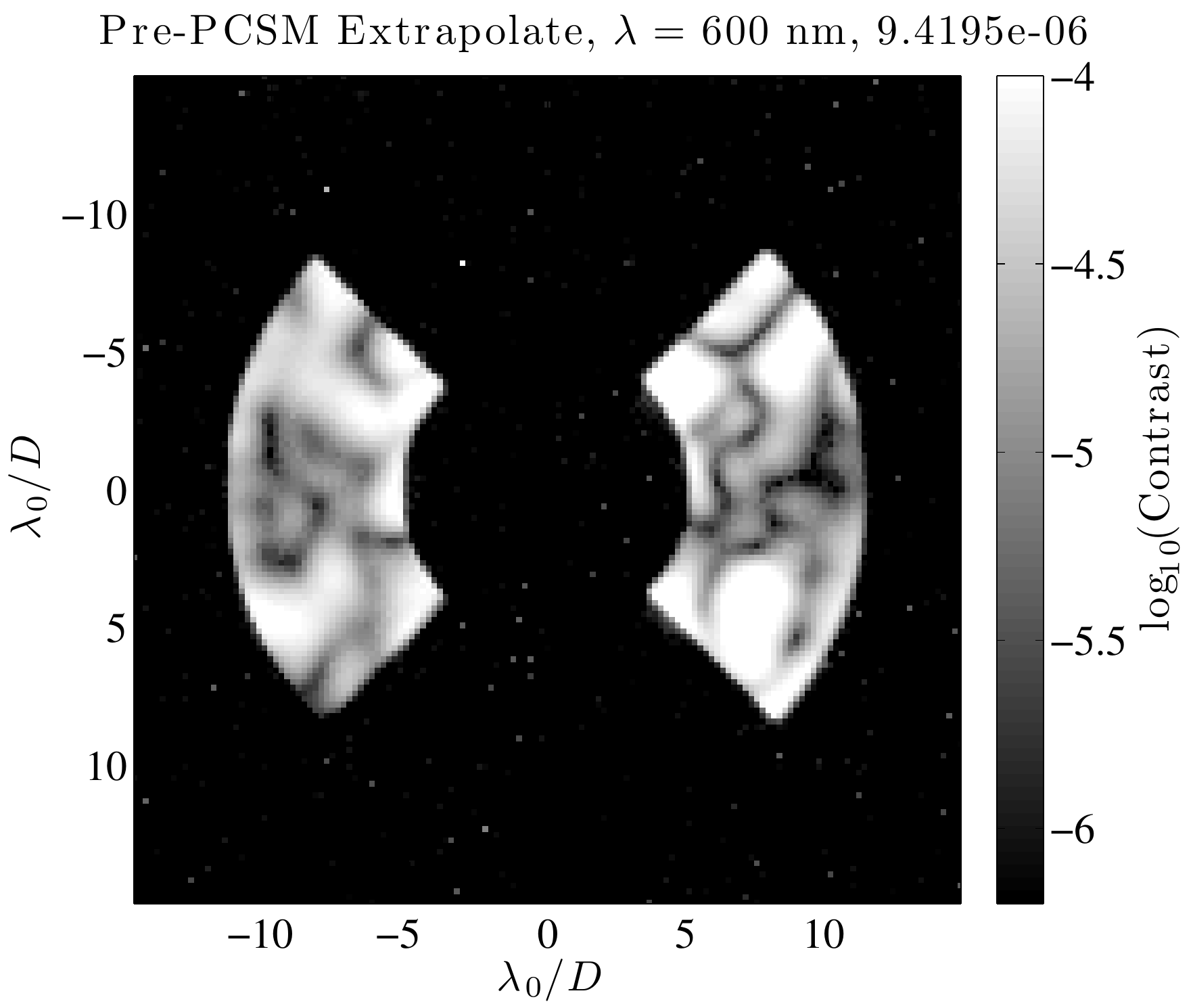}\label{fig:pre_600}}
\subfigure[\bf{$\mathbf{\lambda_0 = 620}$ nm}]{\includegraphics[width = 0.325\textwidth]{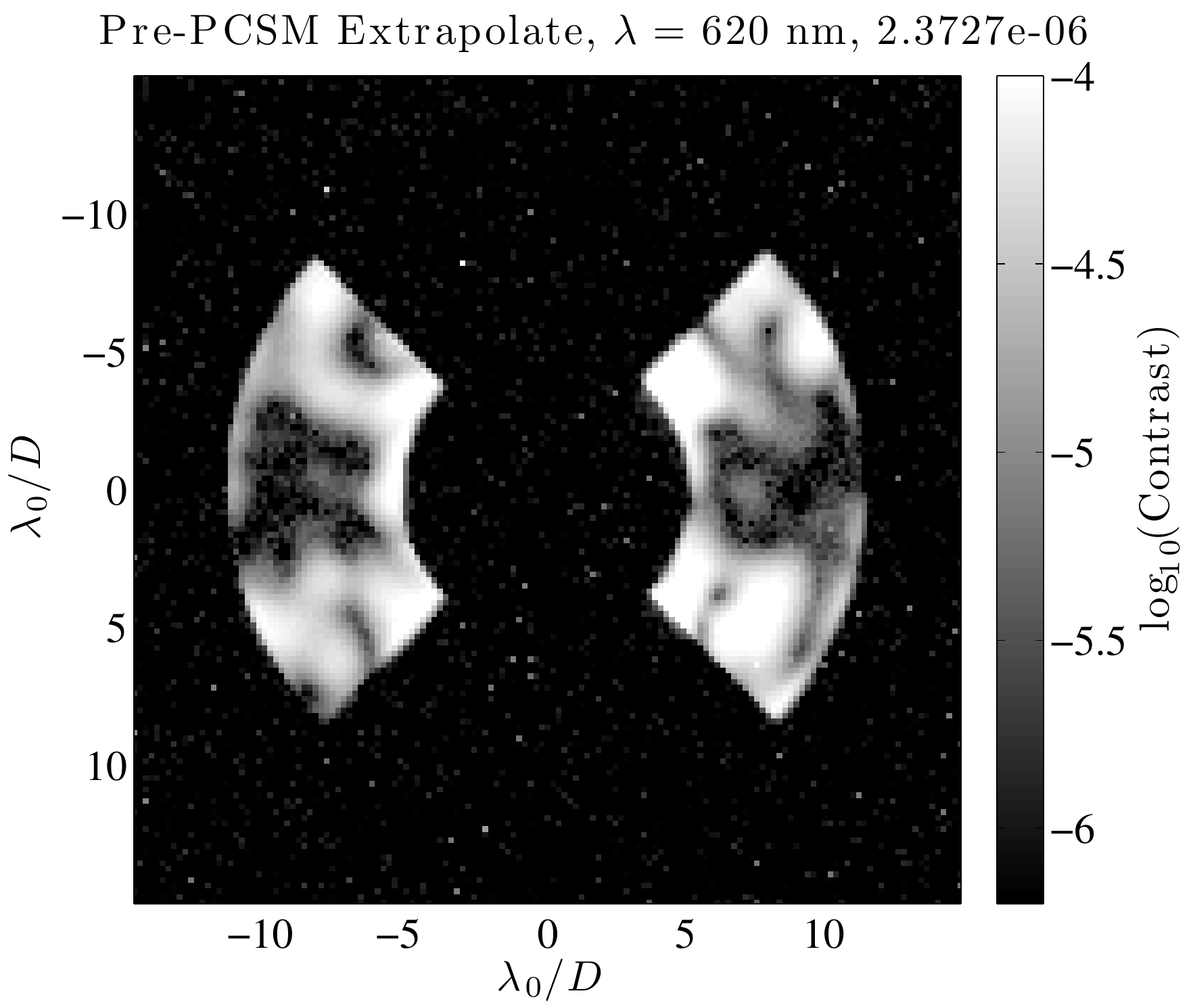}\label{fig:pre_620}}
\subfigure[$\lambda = 650$ nm]{\includegraphics[width = 0.325\textwidth]{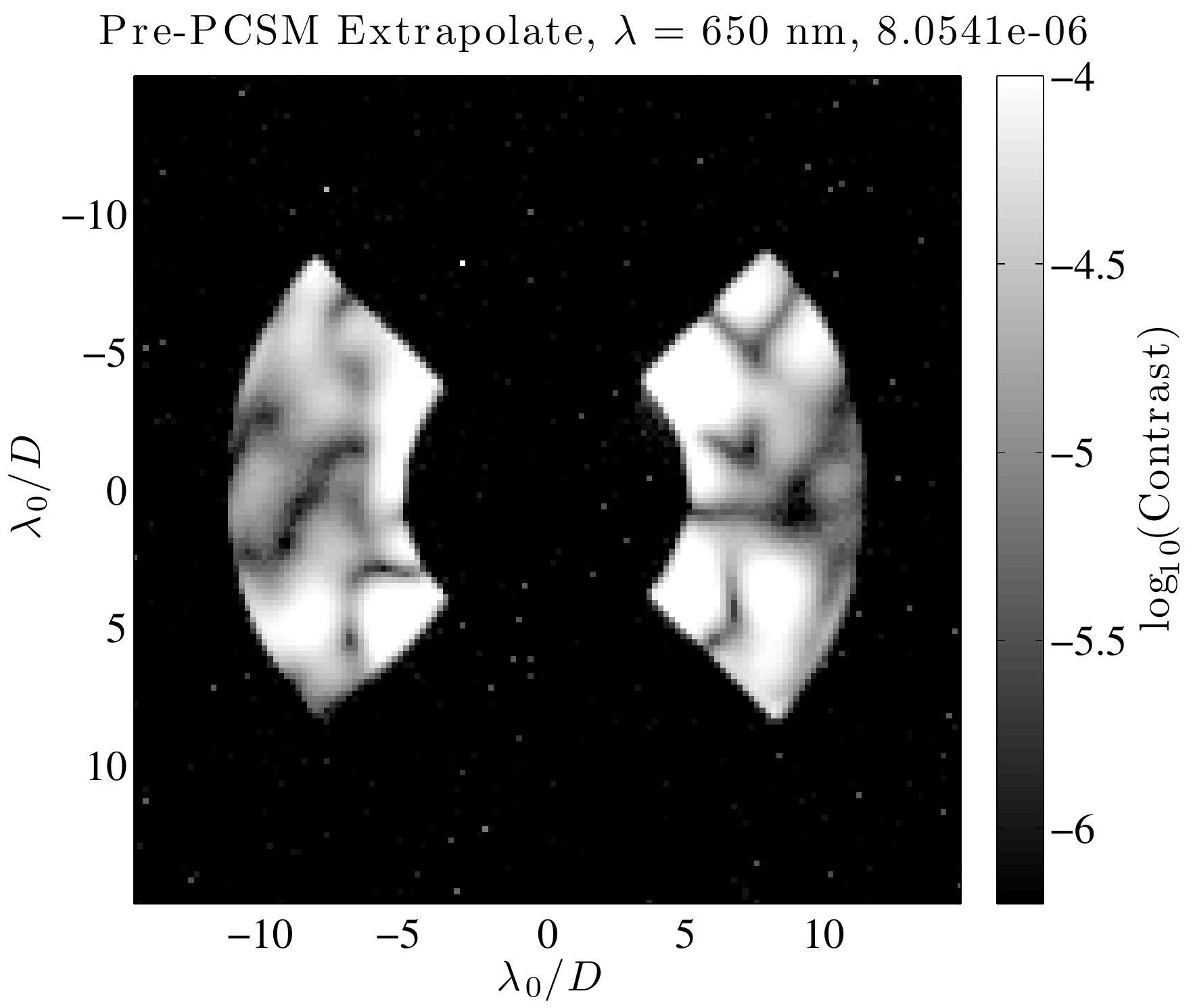}\label{fig:pre_650}}
\caption[Pre-PCSM Extrapolate Individual Filters]{Pre-PCSM Extrapolate Individual Filters}\label{fig:prefilters}
\end{figure}

Looking at the wavelength performance, we see that even the central wavelength is not suppressed particularly well. The dark holes exhibit a good average contrast, but there is a lot of contrast variance within them and their edges are not well defined. Additionally, we see a rapid degradation of the dark hole field as a function of wavelength to the point where it is virtually indistinguishable when we reach the bounding wavelengths in the optimization at $600$ and $650$. Compared to a typical monochromatic experiment, these images depict an abnormally high amount of structure in the dark hole and appear to be highly sensitive to variance in low to mid-spatial frequency aberrations. The chromatic dependence of these errors, particularly at the shorter wavelengths, indicates that the $633$ nm single mode fiber is inadequate for the broadband experiments. This is a result of the multimode output (primarily TEM01 and TEM10 modes) at shorter wavelengths and the high degree of attenuation at longer wavelengths.
\subsection{Photonic Crystal Single Mode Fiber Upgrade}\label{sub:pcsm}
Given the non-single mode nature of the output beam at shorter wavelengths (and our sensitivity to such aberrations), the poor coupling efficiency, and high attentuation of the $633$ nm single mode fiber at longer wavelengths we chose to upgrade the fiber delivery to a Koheras Photonic Crystal continuously Single Mode (PCSM) fiber. We chose the LMA-5 fiber option, which fully spans the bandwidth we operate over, and has the smallest available core diameter (5 $\mu$m). This providesthe smallest mode field diameter available, $4.5$ -- $4.7 \pm 0.5$ $\mu$m, providing a numerical aperture (NA) of ${\approx}0.1$ -- $0.14$ across the visible spectrum (NA being the sine of the divergence half-angle). This is comparable to the $4.3$ -- $4.6$ $\mu$m mode field diameter and NA $0.10$ -- $0.14$ of a single mode $633$ nm fiber between $633$ and $680$ nm. Overall, the PCSM fiber has a lower level of attentuation, is continuously single mode, and roughly matches the beam divergence angle expected from the original fiber, which we have found in the past to well approximate a point source. Since the field from a star is effectively planar, our ability to provide single-mode light at all wavelengths allows us to more accurately demonstrate the controller under conditions true to a real observation.
\begin{figure}[h!]
\centering
\subfigure[]{\includegraphics[width = 0.38\textwidth]{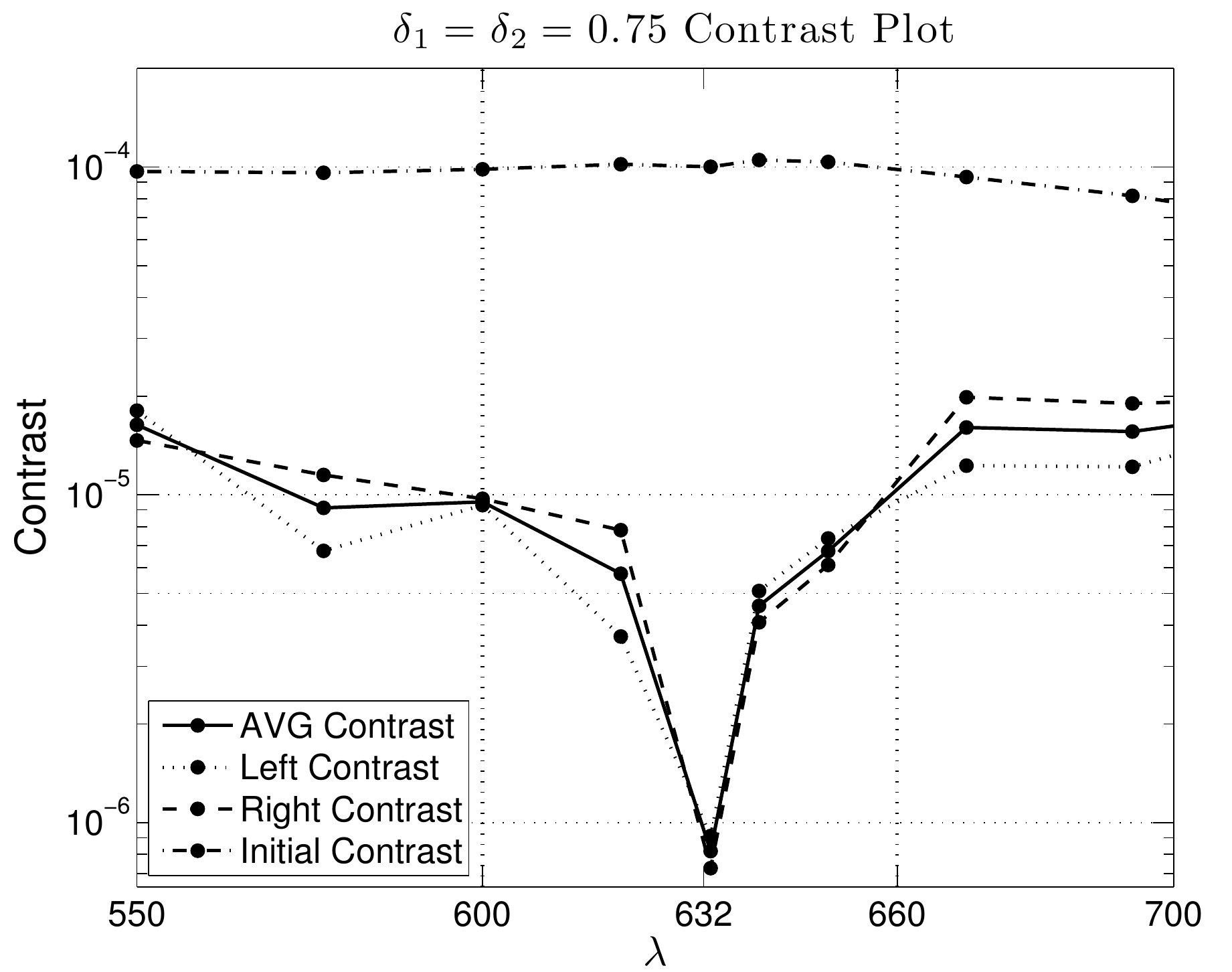}\label{fig:contrastextrap}}
\subfigure[]{\includegraphics[width = 0.35\textwidth]{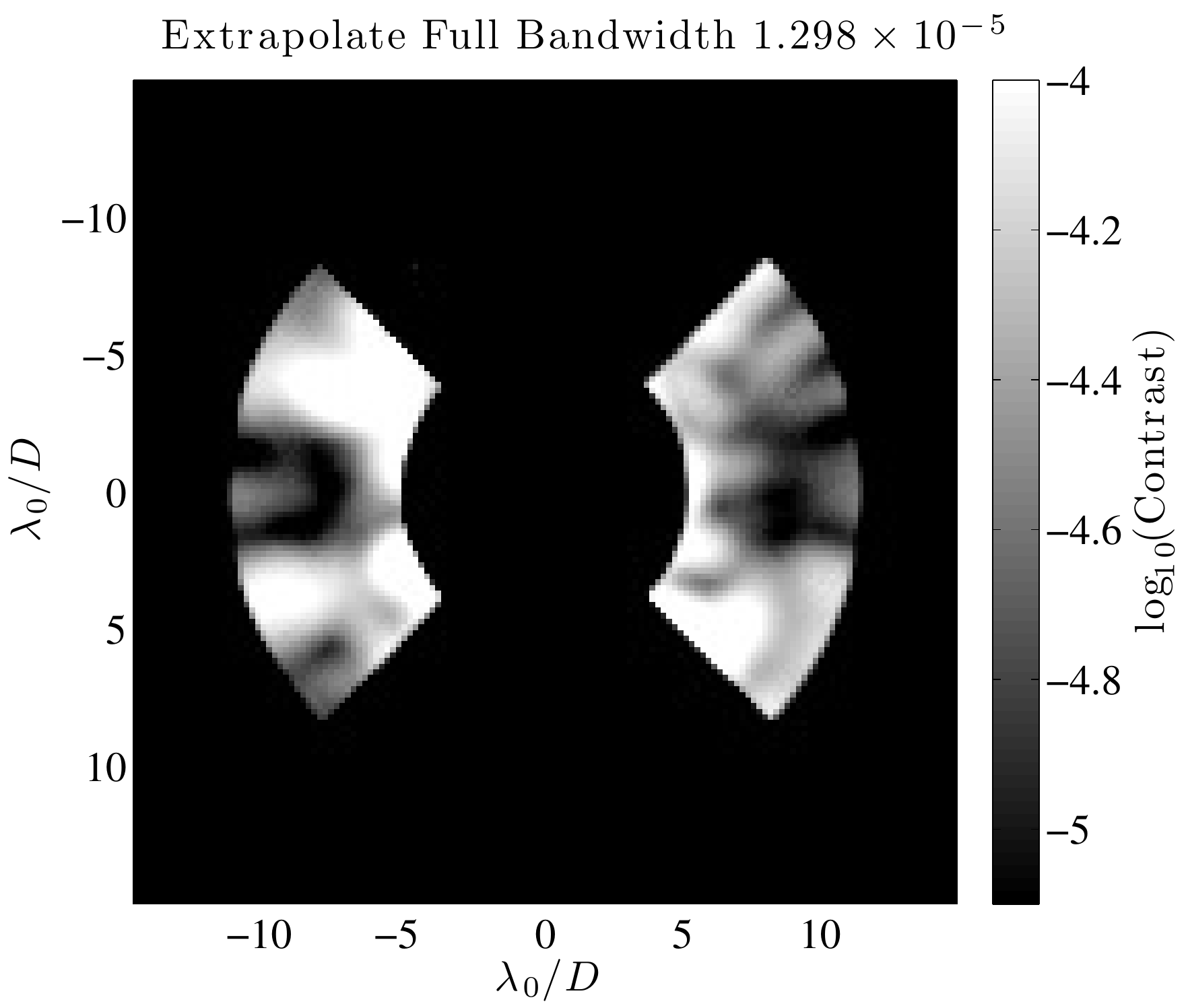}\label{fig:broadextrap}}\\
\subfigure[]{\includegraphics[width = 0.35\textwidth]{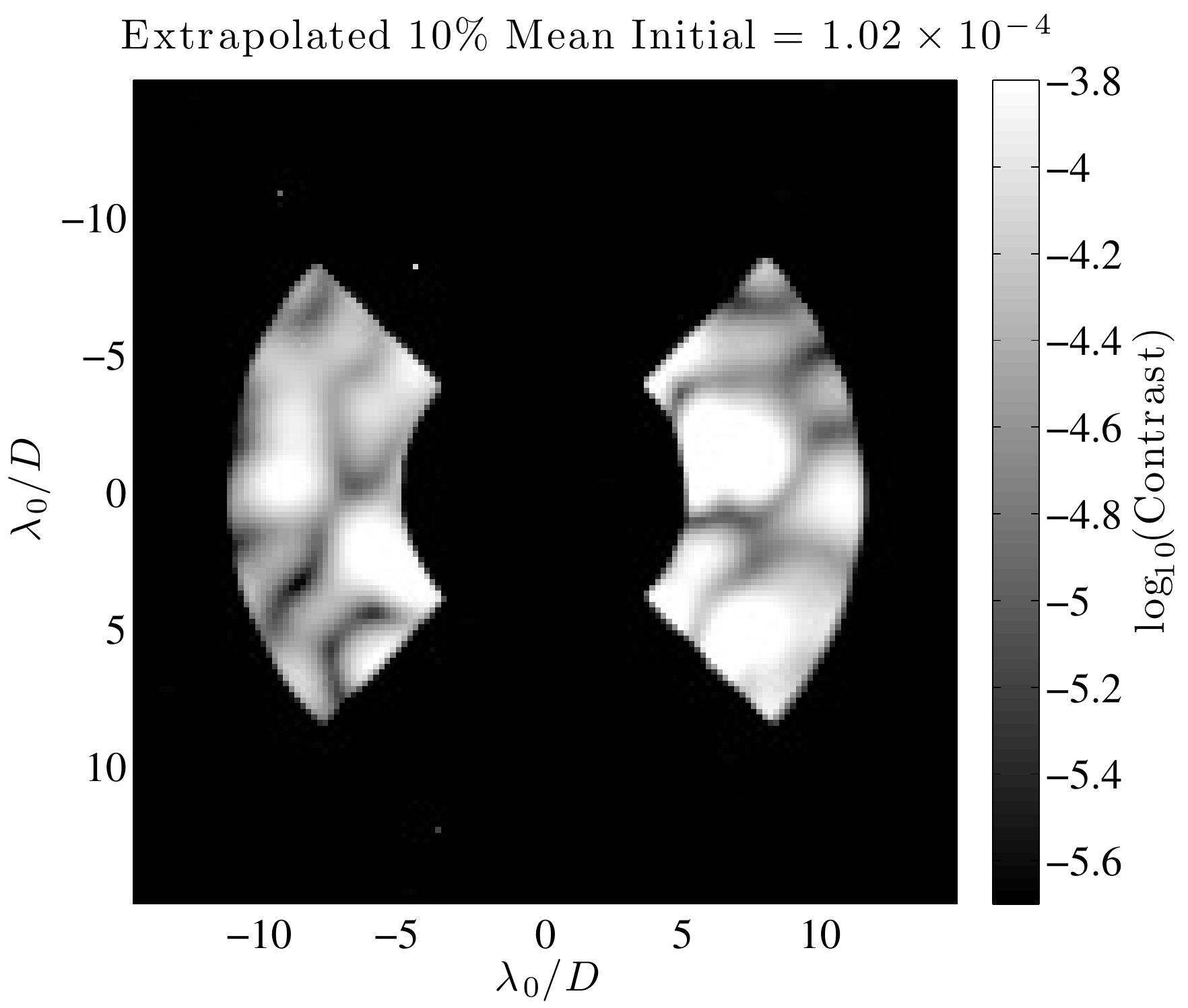}\label{fig:initextrap}}
\subfigure[]{\includegraphics[width = 0.35\textwidth]{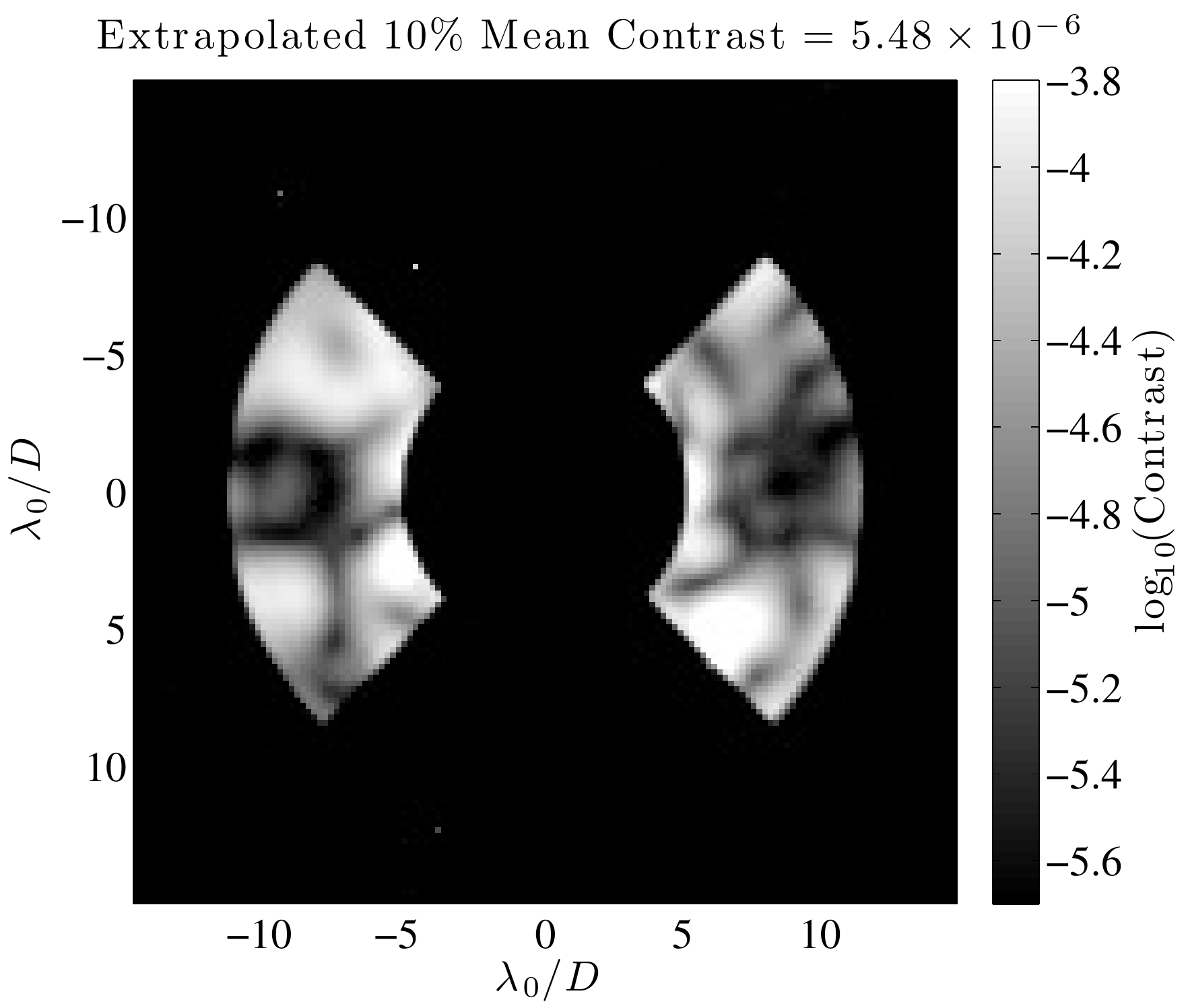}\label{fig:finextrap}}
\caption[Extrapolated Results]{Extrapolated results }\label{fig:extrapresults}
\end{figure}

Fig.~\ref{fig:extrapresults} shows the overall results of applying the same extrapolation technique after the new fiber had been installed. Fig.~\ref{fig:contrastextrap} shows marked contrast improvement at all wavelengths, the out of band wavelengths improving on the order of $3\e{-5}$. As we would have expected, the shorter wavelengths improved more than the longer wavelengths because their output no longer contains higher TEM modes. Comparing Fig.~\ref{fig:broadextrap} to Fig.~\ref{fig:prebroadextrap} we also see that we have a slight improvement in the inner working angle of the dark hole, which is consistent with the fact that we eliminated very low order modes, such as TEM01 and TEM10, by upgrading to the new fiber. Very little of the energy in Fig.~\ref{fig:prefinextrap} is (intentionally) below the cutoff wavelength for single mode output of the $633$ nm SM fiber, which is why the IWA improvement is not as evident when comparing to Fig.~\ref{fig:finextrap}.
\begin{figure}[h!]
\centering
\subfigure[$\lambda = 600$ nm]{\includegraphics[width = 0.325\textwidth]{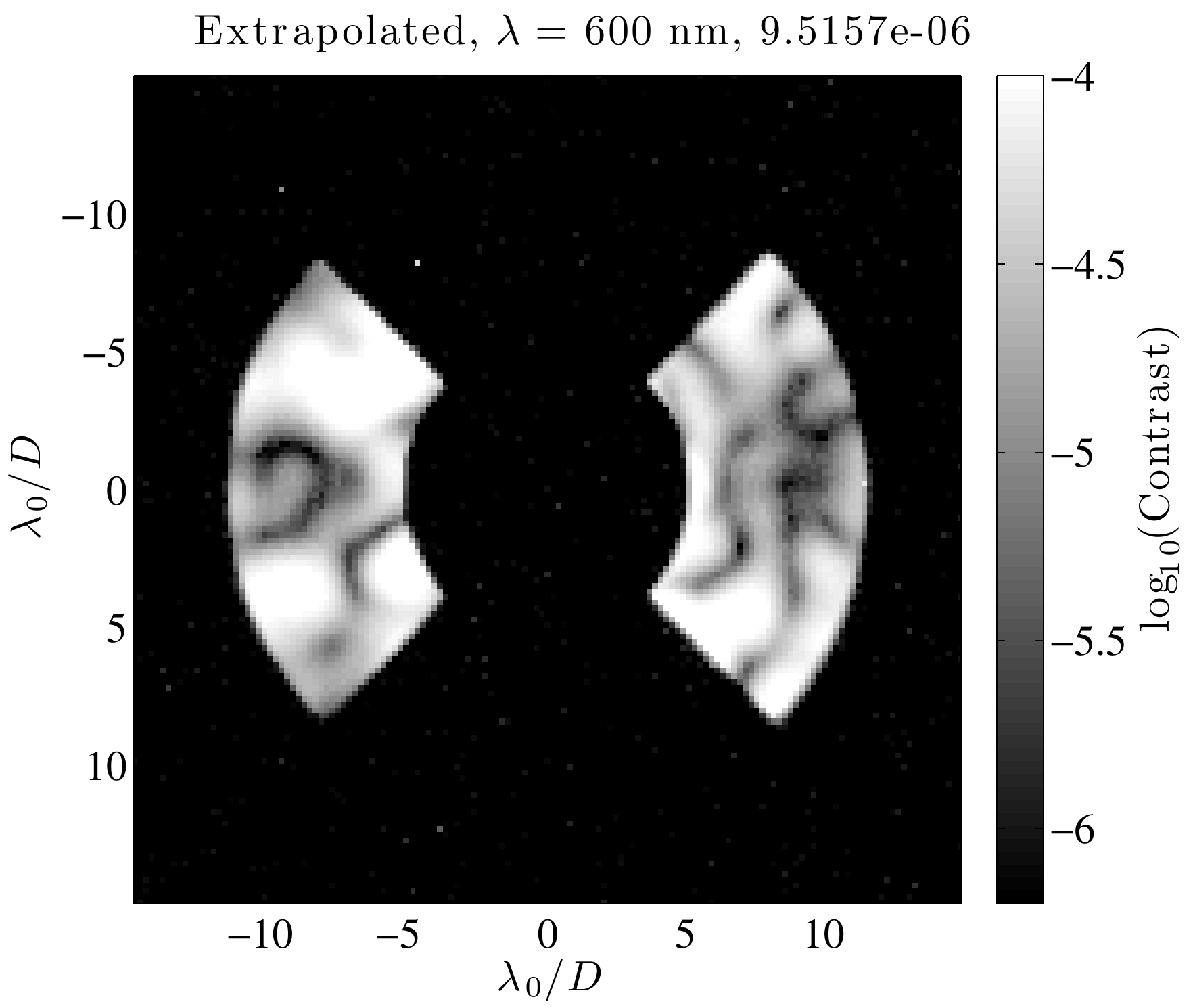}\label{fig:extrap_600}}
\subfigure[\bf{$\mathbf{\lambda_0 = 633}$ nm}]{\includegraphics[width = 0.325\textwidth]{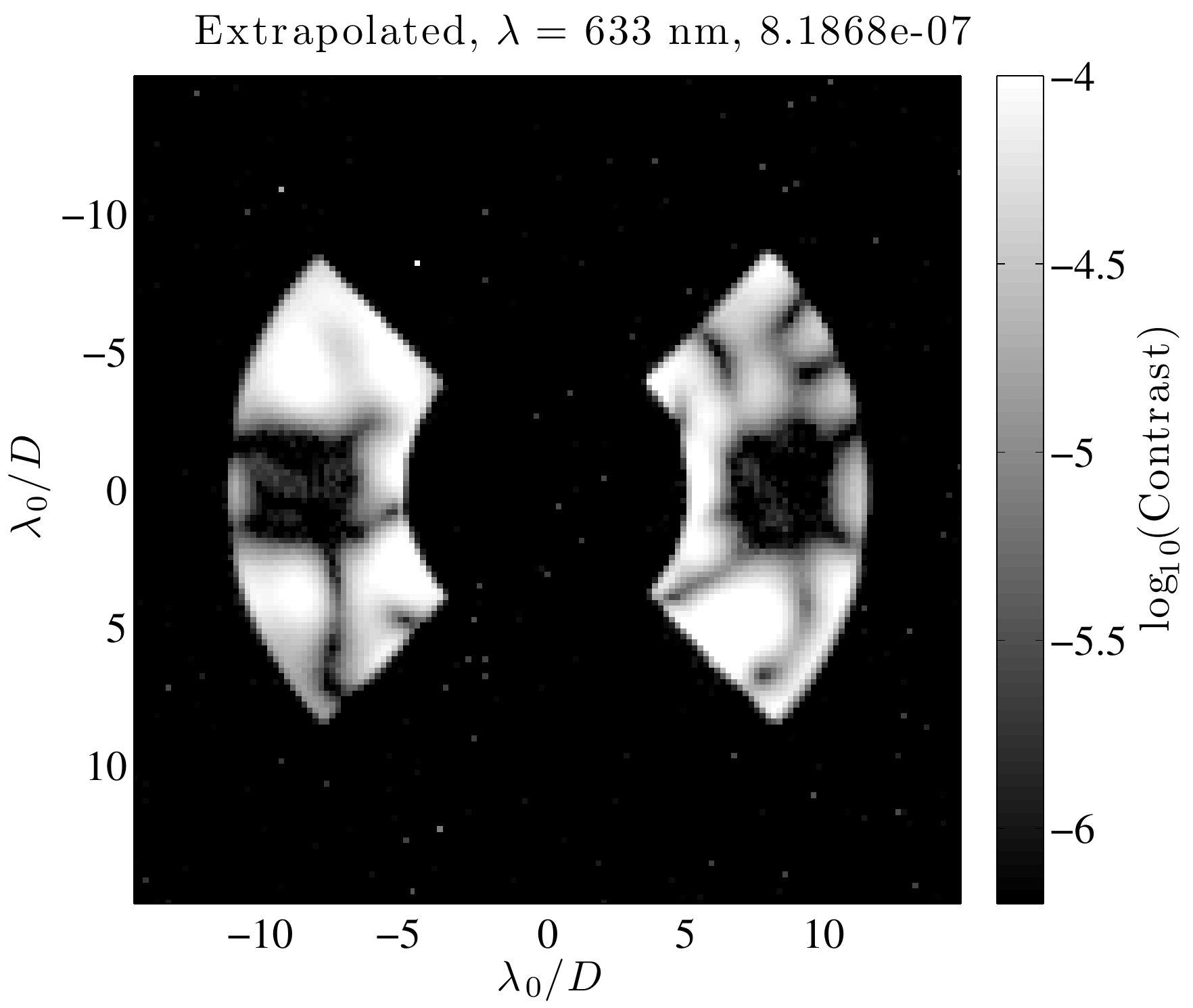}\label{fig:extrap_633}}
\subfigure[$\lambda = 650$ nm]{\includegraphics[width = 0.325\textwidth]{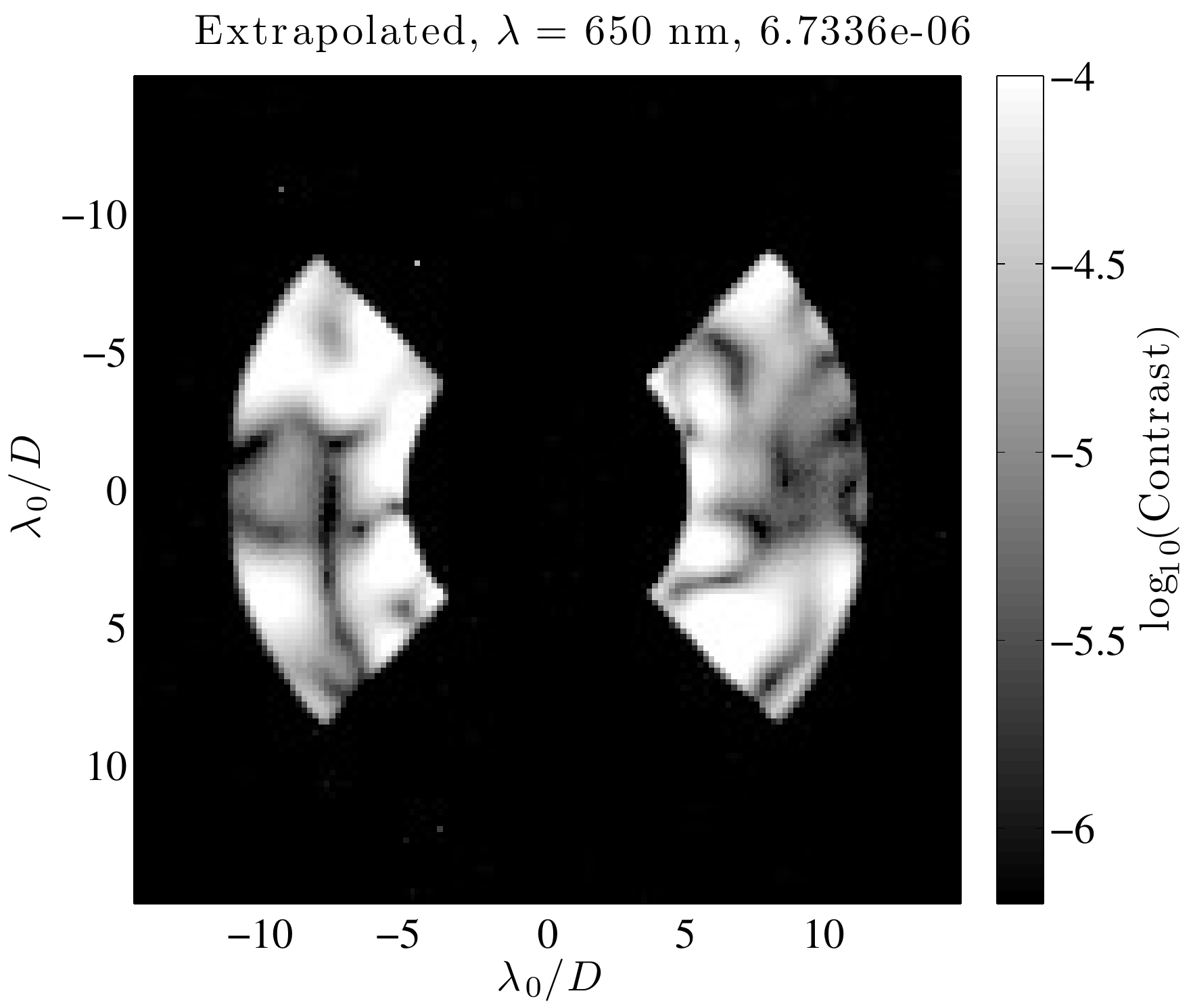}\label{fig:extrap_650}}
\caption[Extrapolated Estimate Individual Filters]{Extrapolated Estimate Individual Filters}\label{fig:extrapfilters}
\end{figure}

Looking at the progression of the final dark hole in wavelength, Fig.~\ref{fig:extrapfilters}, we see that the central wavelength is deeply suppressed  while the intensity of the dark hole raises rapidly. For the filters inside the $\approx10\%$ optimization bandwidth, we see that the contrast degradation is a result of small scale aberrations growing in intensity. Outside of these wavelengths the dark hole degrades rapidly to the point that it is not distinguishable in the $550$ and $740$ nm images. While the average contrast does degrade from the slight shift in the dark hole location, it is also due to speckles within the dark hole increasing in intensity. This indicates that we are somewhat limited by the accuracy of our extrapolation, which tends to introduce fine structure into the dark hole. Note, however, that when comparing Fig.~\ref{fig:extrap_633} with Fig.~\ref{fig:pre_620} we see that the fine structure at the central wavelength is gone. This is entirely due to the fiber upgrade, since no other modifications were made to the experiment.

The accuracy of the functional relationship established in Eq.~\ref{simplepupil} will ultimately bound the achievable bandwidth; therefore, as a metric, these results are also compared to estimating each wavelength separately. Improving this functional relationship requires that we establish a higher order relationship of the electric field that captures more of the system model. For the time being, we compare the performance of the simplest (and fastest) extrapolation technique we may physically motivate to multiple estimates, which will be slower but presumably more accurate at longer wavelengths.
\begin{figure}[h!]
\centering
\subfigure[]{\includegraphics[width = 0.38\textwidth]{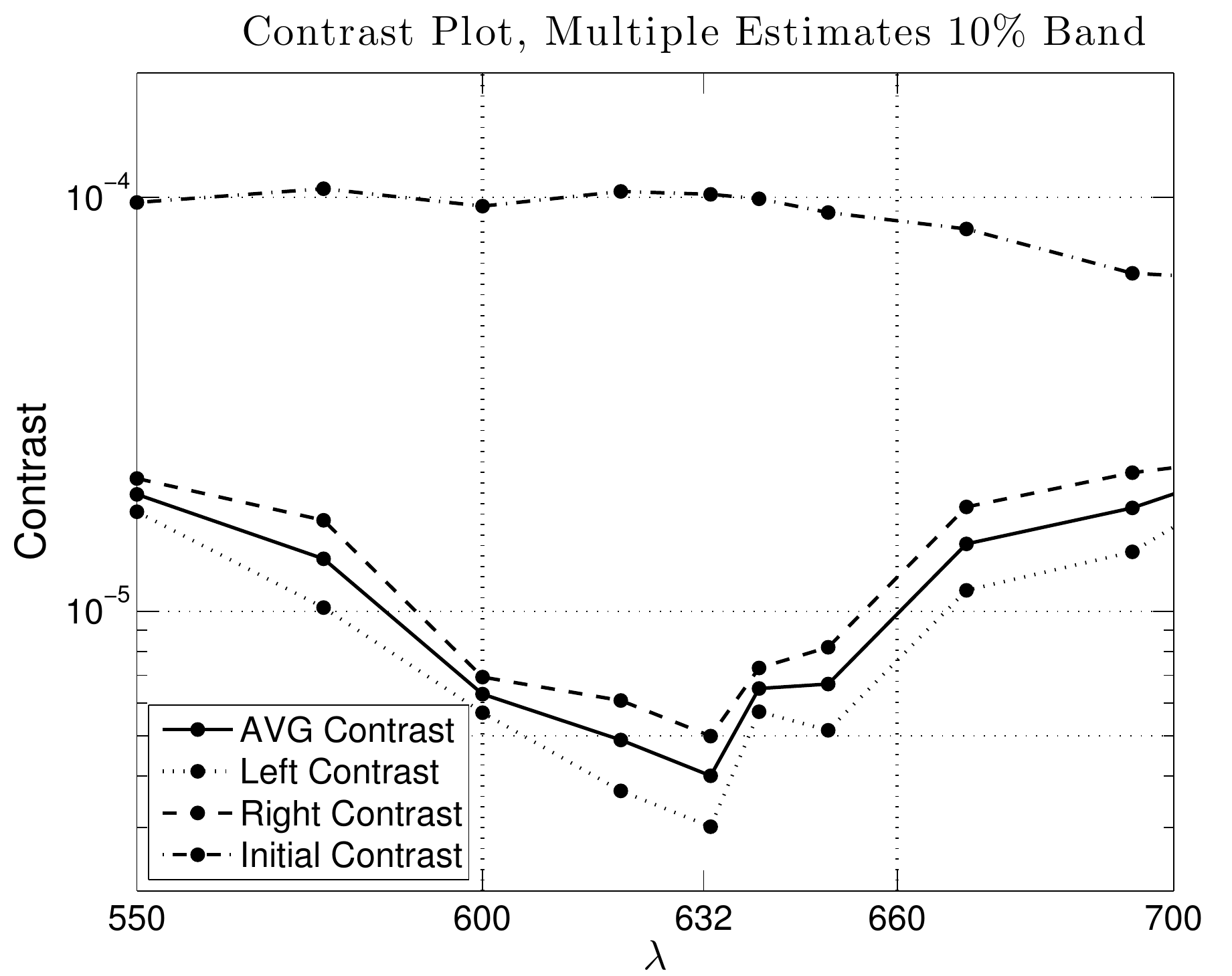}\label{fig:contrastdir}}
\subfigure[]{\includegraphics[width = 0.35\textwidth]{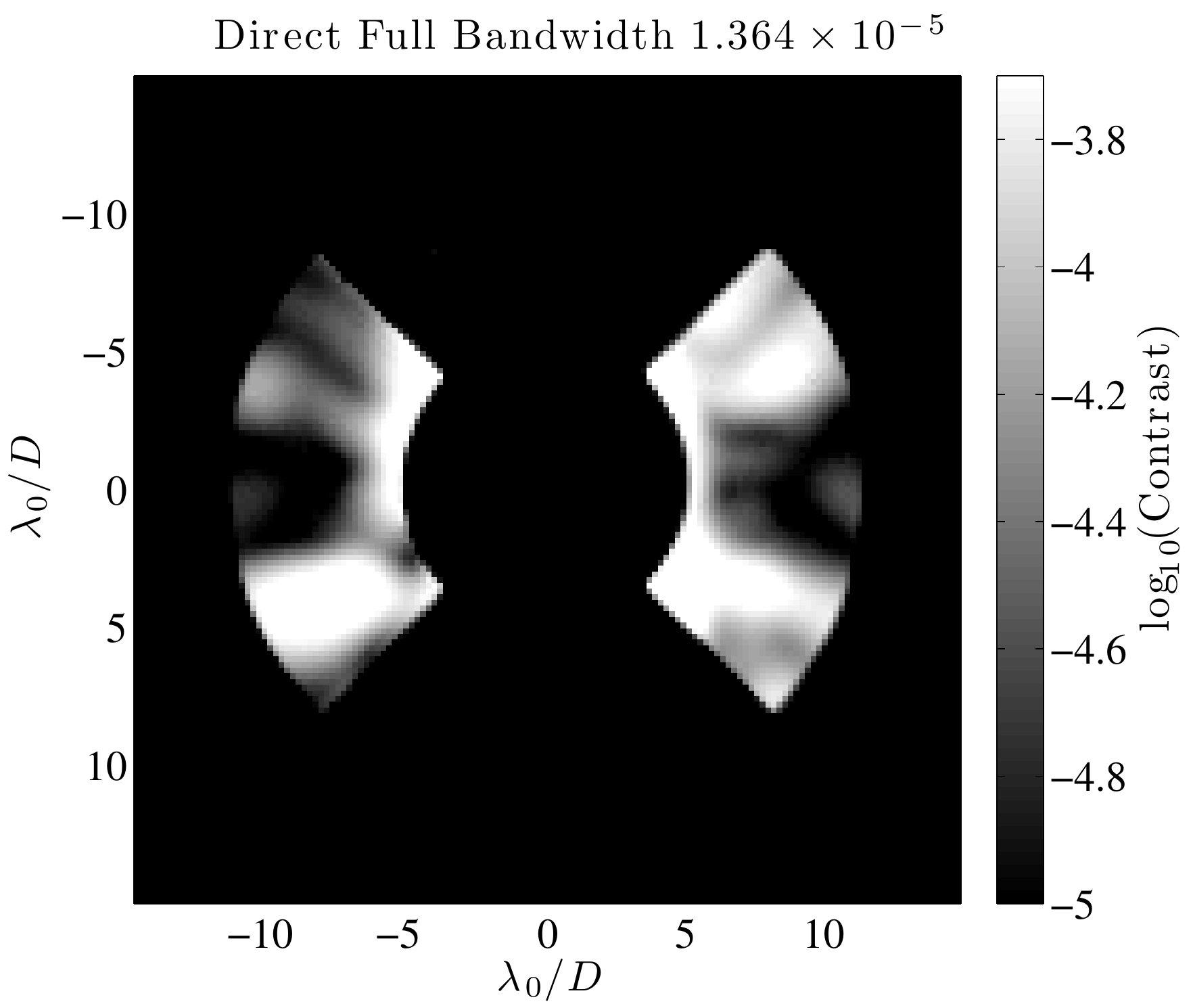}\label{fig:broaddir}}\\
\subfigure[]{\includegraphics[width = 0.35\textwidth]{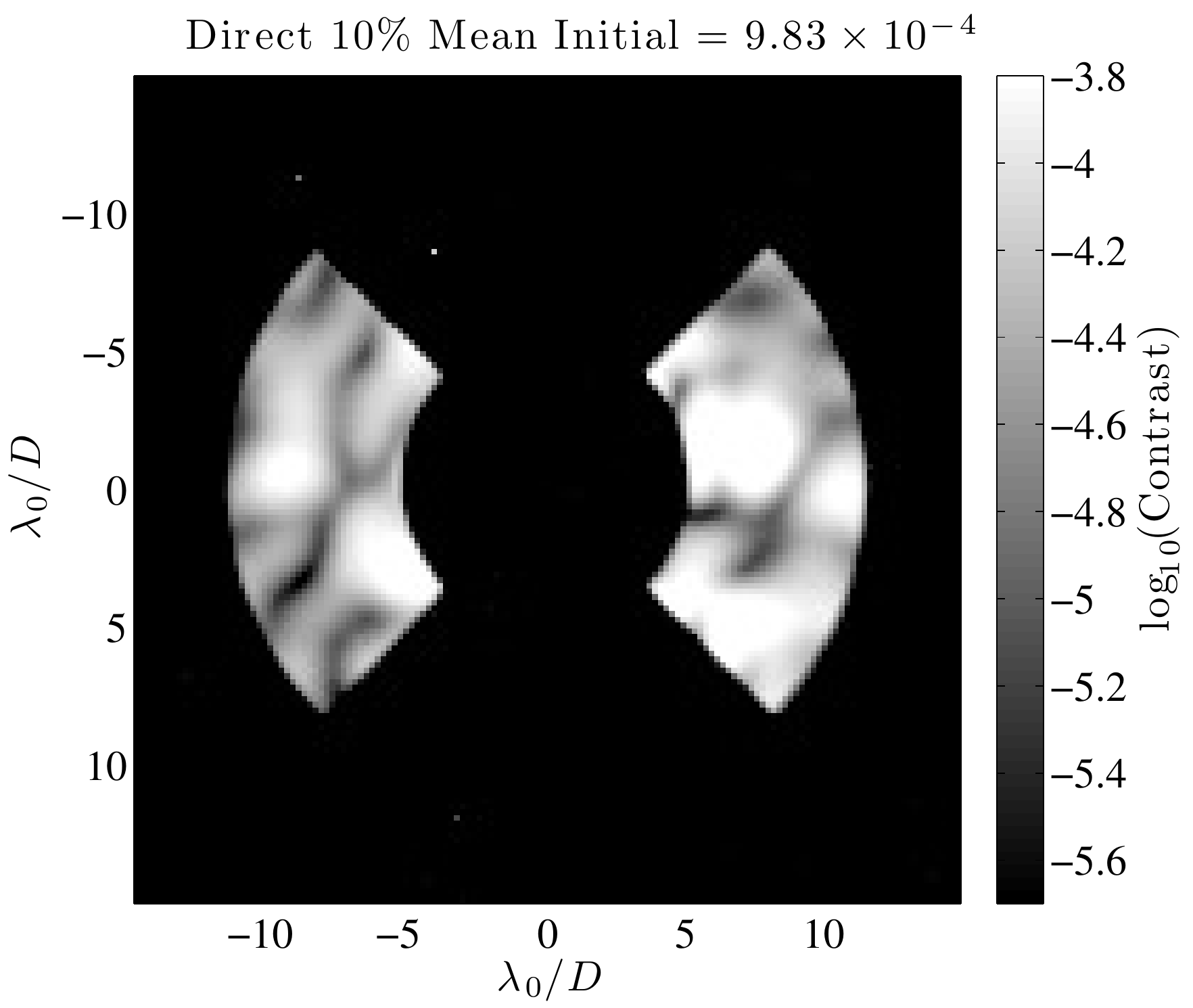}\label{fig:initdir}}
\subfigure[]{\includegraphics[width = 0.35\textwidth]{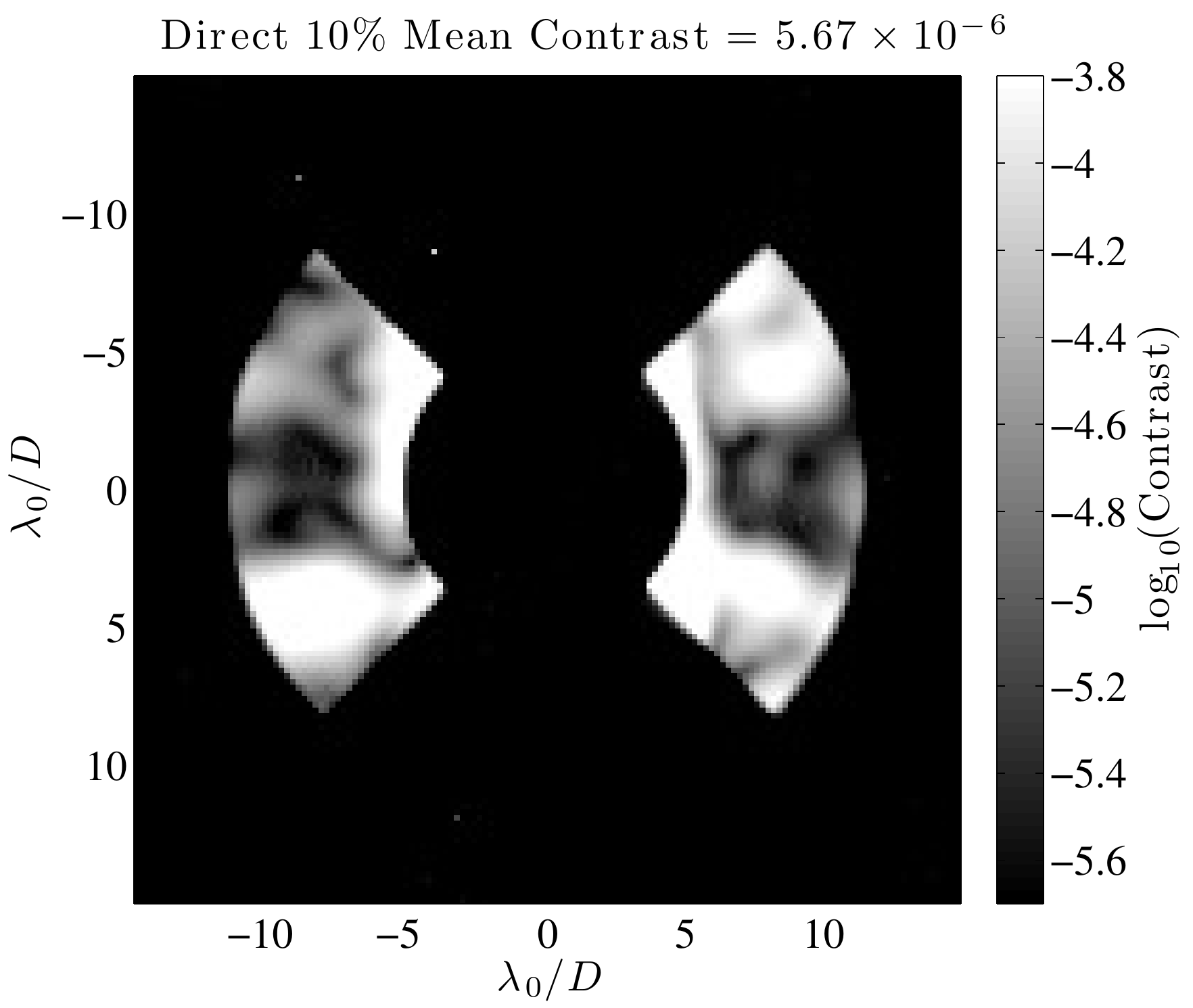}\label{fig:findir}}
\caption[Direct Estimate Results]{Direct Estimate results }\label{fig:direct}
\end{figure}

Fig.~\ref{fig:direct} shows the overall performance of multiple estimates vs. single estimates. When estimating each wavelength separately the contrast reaches $5.67\times10^{-6}$ in a $\sim 10\%$ band (Fig.~\ref{fig:findir}) and $1.364\times10^{-5}$ over the full bandwidth (Fig.~\ref{fig:broaddir}). There is not improvement compared to the $5.67\times10^{-6}$ contrast achieved  in the $10\%$ band and $1.298\e{-5}$ contrast over the full spectrum using the estimate extrapolation technique. Shaklan et al. \cite{shaklan2006terrestrial} show that the ultimate achievable contrast is a function of the correction bandwidth. This limitation is from propagation induced amplitude distributions in the field from surface figure errors on the optics, and the fact that we have a finite controllable bandwidth using two DMs in series (or a Michelson configuration). If we assume that the DM surfaces are the worst figures in our system and apply this to the derivation in Shaklan et al. \cite{shaklan2006terrestrial}, the HCIL optical system should be capable of reaching at least $1\e{-6}$ over a $20\%$ bandwidth, indicating that neither the extrapolation nor the direct estimation methods have reached the fundamental limitations of this optical system (Figs. \ref{fig:contrastextrap}, \ref{fig:contrastdir}) and these results are largely limited by higher sensitivity to estimation error and system stability.
\begin{figure}[h!]
\centering
\subfigure[$\lambda = 600$ nm]{\includegraphics[width = 0.325\textwidth]{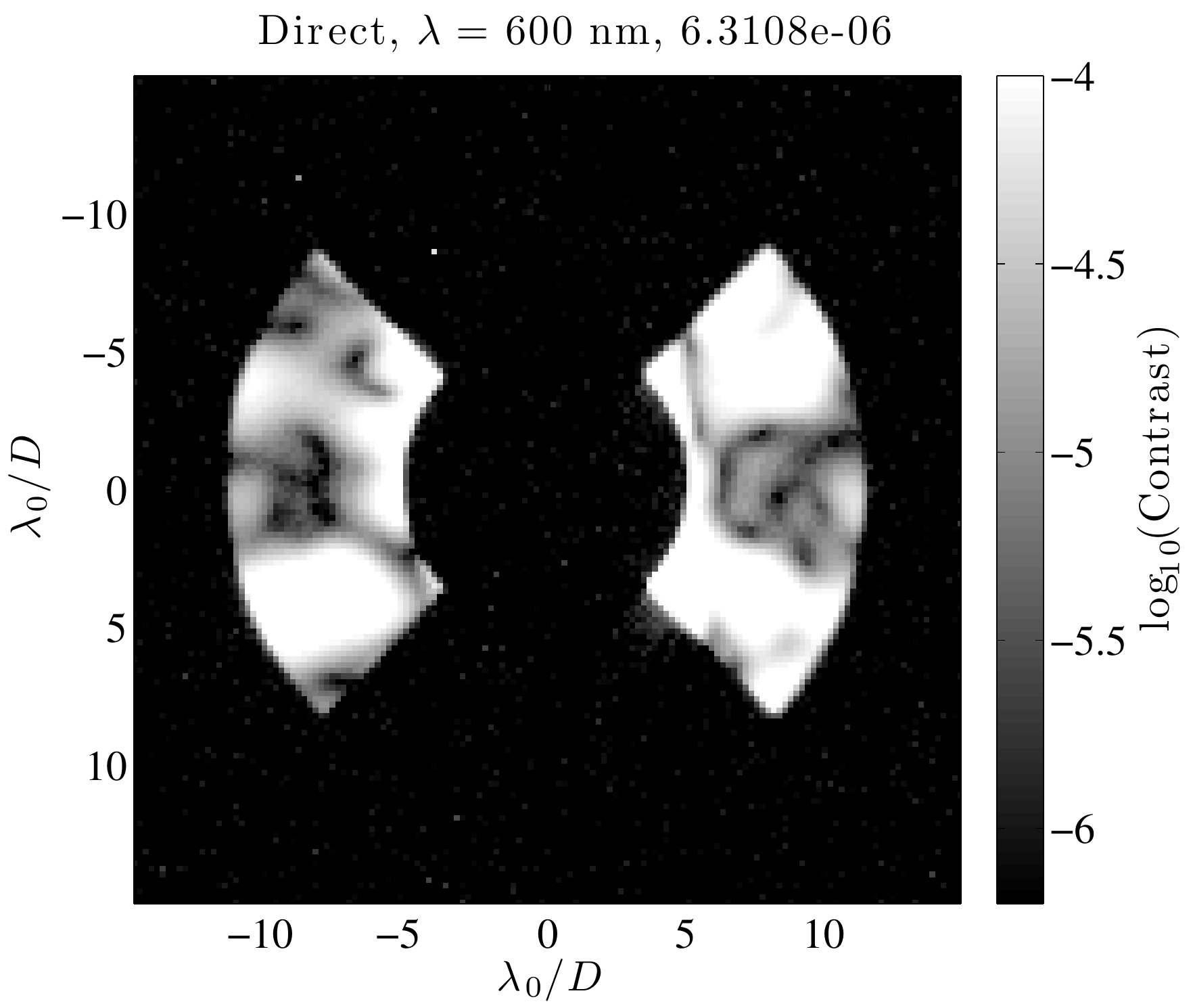}\label{fig:direct_600}}
\subfigure[\bf{$\mathbf{\lambda_0 = 633}$ nm}]{\includegraphics[width = 0.325\textwidth]{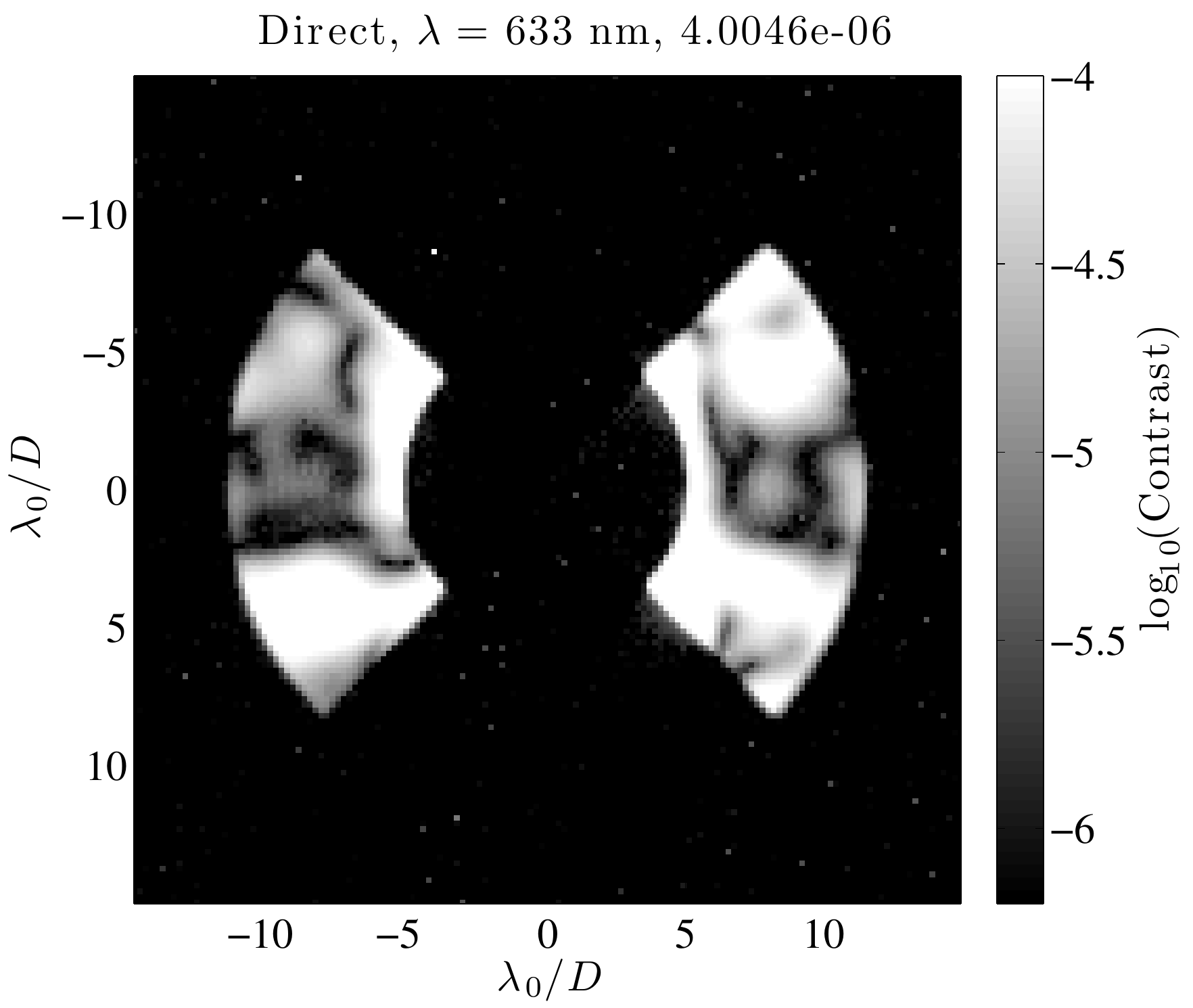}\label{fig:direct_633}}
\subfigure[$\lambda = 650$ nm]{\includegraphics[width = 0.325\textwidth]{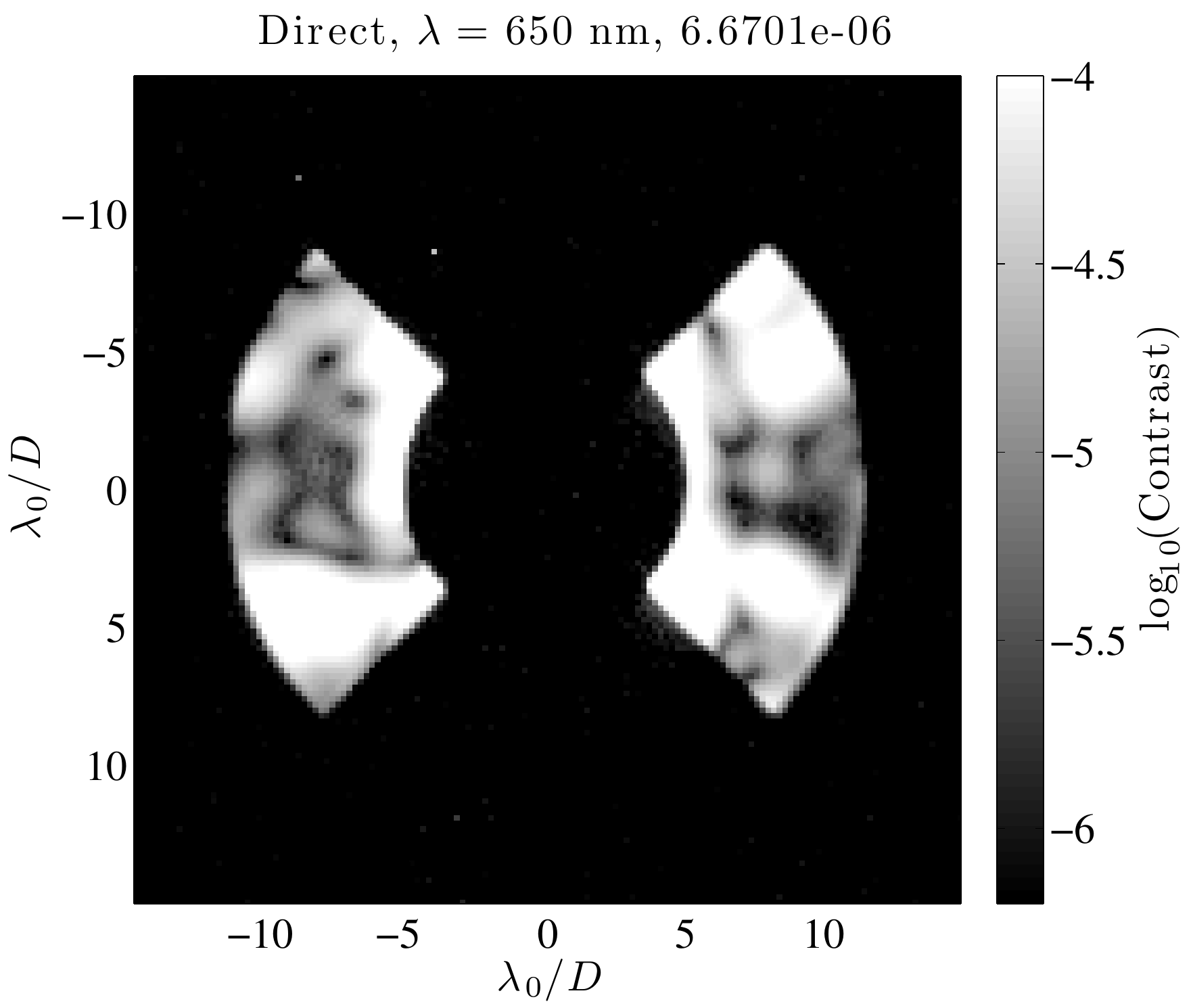}\label{fig:direct_650}}
\caption[Direct Estimate Individual Filters]{Direct Estimate Individual Filters}\label{fig:directfilters}
\end{figure}

Comparing the contrast as a function of wavelength in Fig.~\ref{fig:contrastextrap} and Fig.~\ref{fig:contrastdir}, the bandwidth has been suppressed much more uniformly when multiple estimates are used in lieu of the extrapolation technique.  Since the bounding wavelengths were only slightly underweighted in the optimization ($\mu = 0.75$) we expected a relatively uniform suppression as in Fig.~\ref{fig:contrastdir}. This indicates that the accuracy of the extrapolation was the limiting factor in allowing the controller to evenly suppress the bandwidth. However, the ultimate contrast of the central wavelength is not nearly as low in the direct estimate as it was when applying multiple estimates. Comparing  the dark holes at the central wavelength using estimate extrapolation (Fig.~\ref{fig:extrap_633}), we see that the dark hole using direct measurements (Fig.~\ref{fig:direct_633}) exhibits much more residual structure. However, Fig.~\ref{fig:direct_600}--Fig.~\ref{fig:direct_650} show that the region bounding the corrected area persists better than the dark hole in the extrapolation case, Figs.~\ref{fig:extrap_600}--\ref{fig:extrap_650}. 

We believe that our current achievable contrast in broadband light is limited by the stability of the laboratory. For direct estimation we required three individual estimates to achieve the results shown in Fig.~\ref{fig:direct}. Thus, the estimation step took roughly three times longer than in the extrapolation case. The low power of the filtered broadband light requires exposure times of ${\approx}40$ seconds, as opposed to ${\approx}100-200$ ms in the monochromatic experiment. With $8$ exposures required per estimate using the batch process estimator means that the estimation step went from ${\approx}5$ to ${\approx}15$ minutes per iteration. The HCIL is only stable to ${\approx} 2-5\e{-7}$ over such a long period (independent of power fluctuations). Thus, the extrapolation method reached the limit of system variance over a $5$ minute interval at the central wavelength, but at the cost of less accurate estimates over the bandwidth due to an innaccurate extrapolation. On the other hand, the longer time frame required to take multiple estimates meant that we compromised the stability of the experiment but we were able to more evenly suppress the field over the bandwidth. As a result, we cannot experimentally verify that we have reached a fundamental limit in the laboratory without getting more laser power or improving system stability. The sensitivity of the correction algorithm to laboratory stability demonstrates the power of the extrapolation technique. To take full advantage of an observatory's stability, we clearly want to reduce the time required to produce estimates of the electric field over the optimization bandwidth. Furthermore, the advantage of establishing an augmented cost function and using extrapolated wavelengths is that it automatically extends the optimal estimator developed in Groff et al.\cite{groff2011designing} to broadband light because this method only requires a single monochromatic estimate. It is therefore worthwhile to continue pursuing more accurate and sophisticated extrapolation techniques. The most promising direction we currently see is to augment the Kalman filter to include the extrapolation. This potentially allows us to produce estimates of multiple wavelengths using incomplete measurements at every wavelength. Thus, the estimator could not independently produce an estimate at each wavelength but averages uncertainty in the wavelength dependence across all three estimates.

\section{Conclusions}\label{conclusions}
With the new photonic crystal single mode fiber, we have eliminated multimode output of the fiber source. Comparing the most recent results to data collected before the laboratory upgrade, the extrapolation technique was strongly affected by the multimode output at shorter wavelengths. Taking multiple estimates removes some of this sensitivity, but there was still some improvement because the strong wavelength dependence of the input aberrations will limit our ability to achieve simultaneous correction at multiple wavelengths. With the light source upgrade, the two remaining limitations are algorithmic. One is the quality of our DM model.  Since both the estimation and control algorithms incorporate DM surface models and actuator voltage maps, errors in this mapping directly translate into limitations in contrast.  The last source of error is the extent to which the aberrated field over the wavelength band is well represented by the first two terms in a wavelength expansion. Two DMs in series can only correct the $\lambda$ independent and $1/\lambda$ terms \cite{pueyo2007polychromatic}. The theoretical $1\times10^{-6}$ contrast limit in a $20\%$ band in the Princeton HCIL is driven by the uncontrollable nominal surface errors of the DMs that will contribute $1/\lambda^2$ errors due to Fresnel propagation. This points to a stronger requirement on the figure of the DMs to achieve better broadband suppression in symmetric dark holes, and indicates that there would be great value to adding a third DM in series to the optical system since this would make higher order chromatic errors controllable \cite{pueyo2007polychromatic}.

\section*{Acknowledgements}
This work was funded by NASA Grant \# NNX09AB96G and the NASA Earth and Space Science Fellowship.
\bibliographystyle{spiebib}

\end{document}